\input ./style/arxiv-general.cfg
\documentclass[MSNbibl,nameyear,dvips]{arxstspdf}
\makeatletter
   \@ifpackageloaded{graphicx}{}{\usepackage{graphicx}}
\makeatother
\usepackage{flushend}
\usepackage{stfloats}
\usepackage{url,breakurl}

\volume{30}
\issue{3}
\pubyear{2015}
\firstpage{352}
\lastpage{371}
\doi{10.1214/15-STS513}
\docsubty{FLA}

\makeatletter
\newcommand{\citett}[1]{\citeauthor{#1}, \citeyear{#1}}
\def\i{\mathbf{1}}
\newproclaim{remark}{Remark}
\def\cal{\mathcal}

\def\u{\underline}
\def\what{\widehat}
\makeatother

\begin{document}
\begin{frontmatter}

\title{A Closer Look at Testing the ``No-Treatment-Effect''
Hypothesis in a~Comparative Experiment}
\runtitle{A Closer Look...}

\begin{aug}
\author[A]{\fnms{Joseph B.}~\snm{Lang}\corref{}\ead[label=e1]{joseph-lang@uiowa.edu}} 
\runauthor{J.~B. Lang}

\affiliation{University of Iowa}

\address[A]{Joseph B. Lang is Professor,
Department of Statistics and Actuarial Science, University of Iowa,
207 SH, Iowa City, Iowa 52242, USA \printead{e1}.}
\end{aug}

%
\begin{abstract}
Standard tests of the ``no-treatment-effect'' hypothesis for a
comparative experiment include permutation tests, the Wilcoxon rank sum
test, two-sample $t$ tests, and Fisher-type randomization tests.
Practitioners are aware that these procedures test different no-effect
hypotheses and are based on different modeling assumptions. However,
this awareness is not always, or even usually, accompanied by a clear
understanding or appreciation of these differences. Borrowing from the
rich literatures on causality and finite-population sampling theory,
this paper develops a modeling framework that affords answers to
several important questions, including:
exactly what hypothesis is being tested, what model assumptions are
being made, and are there other, perhaps better, approaches to testing
a no-effect hypothesis? The framework lends itself to
clear descriptions of three main inference approaches: process-based,
randomization-based, and selection-based. It also promotes careful
consideration of model assumptions and targets of inference, and
highlights the importance of randomization. Along the way, Fisher-type
randomization tests are compared to permutation tests and a less
well-known Neyman-type randomization test. A simulation study compares
the operating characteristics of the Neyman-type randomization test to
those of the other more familiar tests.
\end{abstract}

%
\end{frontmatter}

\section{Introduction}\label{sec1}

We begin with a simple example of a randomized comparative experiment.
Researchers are interested in determining whether cell phone use while
driving has an impact on reaction times. Toward this end, 64 University
of Utah student volunteers were enlisted to take part in a randomized
comparative experiment (\citett{STRJOH01}). Of the 64 students,
32 were randomized to treatment~1 (operate a driving simulator while
using a cell phone) and 32 were randomized to treatment~2 (operate a
driving simulator without a cell phone). For a summary description of
the data and of the way the two treatments were actually administered,
see Agresti and Franklin (\citeyear{AGRFRA07}, page~446). In the driving simulation, each
student encountered several red lights at random times. Each student's
response was the average time required to stop when a red light was
detected. The 64 responses, in milliseconds, are recorded in Table~\ref{tab1}.

\begin{table*}
\caption{Reaction times (milliseconds)}\label{tab1}
\begin{tabular*}{350pt}{@{\extracolsep{\fill}}ll@{}}
 Cell phone: &636\ 623\ 615\ 672\ 601\ 600\ 542\ 554\ 543\ 520\
609\ 559\ 595\ 565\ 573\ 554\\
&626\ 501\ 574\ 468\ 578\ 560\ 525\ 647\ 456\ 688\ 679\ 960\ 558\ 482\ 527\
536\\
 Control: &557\ 572\ 457\ 489\ 532\ 506\ 648\ 485\ 610\ 444\ 626\
626\ 426\ 585\ 487\ 436\\
&642\ 476\ 586\ 565\ 617\ 528\ 578\ 472\ 485\ 539\ 523\ 479\ 535\ 603\ 512\
449\\
 Generically\ldots\\
 Treatment 1: &$y_{1,1}, y_{1,2},\ldots, y_{1,32}$ \\
 Treatment 2: & $y_{2,1}, y_{2,2},\ldots, y_{2,32}$.
\end{tabular*}
\end{table*}

Is there a cell phone use effect? Generically, is there a \emph{treatment effect}?

Standard tests of the ``no-treatment-effect'' hypothesis include
permutation tests (Pitman, \citeyear{PIT37}, \citeyear{PIT38}), the Wilcoxon rank sum test
(\citett{WIL45}), two-sample $t$ tests (cf. \citett{WEL38}), and Fisher-type
randomization tests (\citett{EDEYAT33,FIS35}; see also the
history in \citett{Dav08}). Most practitioners are aware that these
procedures test different ``no-effect'' hypotheses and are based on
different modeling assumptions. However, this awareness is not always,
or even usually, accompanied by a clear understanding or appreciation
of these differences. This paper looks at each of these testing
approaches and addresses the all important questions, exactly what
hypothesis is being tested and what model assumptions are being made?
Along the way, we will have to confront several other questions such
as, how is the definition of \emph{treatment effect} operationalized,
what is the actual target of inference, what is the role of
randomization, and are there other, perhaps better, approaches to
testing a no-effect hypothesis?

To address these questions, we draw on ideas from the rich literature
on causal analysis. In particular, we employ the useful concept of
``potential variables.'' Although the idea of potential variables can be
traced back to Neyman (\citeyear{Ney90}), Rubin, beginning with a series of papers
on causal models in the 1970s (see \citett{RUB10}, and references therein)
is usually credited with more explicitly stating the potential variable
model and extending it to both randomized and nonrandomized design
settings, with or without covariates (see Rubin's causal model,
\citett{Hol86}). Between Neyman and Rubin, potential variables were used
by relatively few authors; \citet{WEL37}, Kempthorne (\citeyear{Kem52}, \citeyear{Kem55}), and
Cox (\citeyear{COX58}), Section~5, were among the notable early proponents. Around the
time of and after Rubin, many more authors made important contributions
to the potential variables literature. See, for example, \citet{Cop73},
\citet{Bai81}, \citet{Hol86}, Greenland (\citeyear{GRE91}, \citeyear{GRE00}), \citet{Gad01},
and the references therein.

To be clear, it is not the goal of this paper to summarize the vast
literature on potential variables and causal modeling. (To this end,
see Paul R. Rosenbaum's very informative website and referen\-ces
therein, \surl{www-stat.wharton.upenn/\textasciitilde rosenbap/\\downloadTalks.htm}.) Instead, the first goal is to exploit
the benefits of hindsight to develop a modeling framework that supports
clear descriptions and comparisons of the different testing approaches,
and promotes careful consideration of the model assumptions and targets
of inference. This modeling framework and associated notation draws
clear distinctions between realizations and random variables, and
between observed and unobserved data. It accommodates both treatment
assignment and sampling from populations, and clearly differentiates
between the two. Although the proposed model lends itself to
generalizations in many directions (e.g., more than two treatments,
restricted randomization, etc.), to simplify exposition, we will focus
on the two-treatment comparative experiment setting. This restriction
allows us to more directly highlight the useful features of the
proposed modeling framework.

The second goal of this paper is to address the question of
availability of other testing approaches, besides the four common ones
mentioned above.
Toward this end, we revisit ideas introduced in Neyman (\citeyear{Ney90}). Using
the model structure introduced herein, we describe a less well-known
Neyman-type randomization test, which is qualitatively different than
the Fisher-type randomization test (cf. \citett{WEL37}; Rubin, \citeyear{Rub90}, \citeyear{RUB04},
\citeyear{RUB10}). (Readers with an interest in history are encouraged to read
Neyman et al., \citeyear{NEYetal35}, along with the discussions, to see how Neyman and Fisher
publicly aired their differences of opinions on testing in randomized
design settings.) The Neyman randomization test, which uses a less
restrictive ``no-effect'' hypothesis than Fisher's, is based on a test
statistic with the common form, (estimator minus estimand)/(standard
error of estimator). Neyman, with an eye on interval estimation rather
than testing, derived the standard error with respect to a
randomization distribution using tools from finite-population sampling
theory. In retrospect, Neyman's derivation approach is hardly
surprising given that he ``may be said to have initiated the modern
theory of survey sampling'' (Lehmann, \citeyear{LEH94}) in his landmark paper of
1934 (\citett{NEY34}). Compared to Fisher randomization tests, the \mbox{Neyman}
tests do have their advantages and disadvantages. One disadvantage is
that \mbox{Neyman} tests are approximate, whereas Fisher tests are exact. An
advantage is that the \mbox{Neyman} test can be more powerful than the Fisher
test (see Section~\ref{sec7} below). Another advantage is that, unlike the
Fisher randomization test, the Neyman version can be used to test
hypotheses about a population when units are randomly sampled from the
population and then randomized to treatment levels.

The third and final goal of this paper is to compare the operating
characteristics of the five tests: the permutation test, the Wilcoxon
rank sum test, the two-sample $t$-test, the Fisher randomization test,
and the Neyman randomization test. The penultimate section of the paper
includes a small-scale simulation study of the size and power of these
five tests. Based on these comparisons, we make tentative
recommendations on which test to use in different settings.

\begin{table*}
\caption{Potential values notation}\label{tab2}
\begin{tabular*}{\textwidth}{@{\extracolsep{\fill}}lll@{}}
\hline
Treatment 1: &$y_{1.s_1}\hspace*{-18pt} \mbox{\Large$\times
$},\hspace*{6pt} y_{1.s_2}\hspace*{-18pt} \mbox{\Large$\times$},\hspace*{6pt} y_{1.s_3},
\ldots, y_{1.s_{63}} \hspace*{-18pt}\mbox{\Large$\times
$},\hspace*{6pt} y_{1.s_{64}}$& Only the 32 non-\mbox{\Large$\times$}'ed
out values are observed.\\
Treatment 2:& $y_{2.s_{1}}, y_{2.s_{2}}, y_{2.s_3}\hspace*{-18pt}
\mbox{\Large$\times$}\hspace*{6pt}, \ldots, y_{2.s_{63}},
y_{2.s_{64}} \hspace*{-18pt}\mbox{\Large$\times$}$&
Only the 32 non-\mbox{\Large$\times$}'ed out values are observed.\\
\hline
\end{tabular*}
\tabnotetext[]{}{Here, $\u s = (s_1,\ldots, s_{64})$ is a sample from some
population $\u P = (1,\ldots, N), N \geq64$.}
\end{table*}

The remainder of this paper is organized as follows: Section~\ref{sec2}
introduces potential variables and recasts the data in Table~\ref{tab1} within
this framework. Section~\ref{sec3} introduces a sequential data generation model
that explicitly accommodates both random sampling and randomization.
The components in the three-level sequential model are identified as
the ``process,'' the ``sampling,'' and the ``randomization.'' This model,
along with a useful component-selection notation, leads to an explicit
identification of the observed data and the three main targets of
inference. Section~\ref{sec4} gives candidate definitions of treatment effects
that are based on potential variables; corresponding
no-treatment-effect hypotheses are also given. An overview of the three
main inference approaches---process-based, selection-based, and
randomization-based---is given in Section~\ref{sec5}. Section~\ref{sec6} introduces a
difference statistic that can be used as the basis for tests of the
no-treatment-effect hypotheses. Tests corresponding to each
of the three inference approaches, along with assumptions for their
validity, are described in detail; some of these tests are well known
and some are less well known. Section~\ref{sec7} carries out an analysis of the
cell phone data and includes a small-scale simulation study of the
operating characteristics of the different testing approaches discussed
herein. Finally, Section~\ref{sec8} includes a brief discussion.

\section{What Might Have Been: The Potential Variables Viewpoint}\label{sec2}

Going back to Neyman (\citeyear{Ney90}) and following the lead of \citet{WEL37},
Kempthorne (e.g., \citeyear{Kem55,Kem77}), Cox (\citeyear{COX58}), and Rubin (e.g., \citeyear{Rub05}), we
will view the data as observed values of a sample of ``potential values.''

Consider a population of $N$ units that are, without loss of
generality, identified by the numbers $1$ through $N$; in symbols,
we will let $\u P = (1,\ldots, N)$ represent the unit identifiers for
the population. For convenience, we will also refer to $\u P$ as the
population of units. Let $Y_{t.i}$ be the response for unit $i$ when
exposed to treatment $t$, where $i=1,\ldots, N$ and $t = 1, 2$.
The response variables $Y_{1.i}$ and $Y_{2.i}$ are called potential
variables for reasons made clear in the next paragraph.

The introduction of these potential variables leads to intuitively
appealing definitions of treatment effects that are based on
head-to-head comparisons of $Y_{1.i}$ and $Y_{2.i}$. There is a catch,
however. Although there is the potential to observe either $Y_{1.i}$ or
$Y_{2.i}$, unfortunately, it is not possible to observe both. Strictly
speaking, it is not possible to observe the values of both potential
variables because the same subject cannot be simultaneously exposed to
both treatments. To the potential variable advocates, this is the
``fundamental problem of causal analysis'' (\citett{Hol86}). As an
example, if we observe the value of $Y_{2.i}$, then the value of
$Y_{1.i}$, and hence the difference $Y_{1.i} - Y_{2.i}$, cannot be
observed. In this case, the unobserved value of $Y_{1.i}$ is relegated to
counterfactual status; the value is ``what might have been'' had unit
$i$ been exposed to treatment 1 rather than treatment 2.

The data in Table~\ref{tab1} can be viewed as observed values of a sample of the
potential variable values. Specifically, a sample $\u s$ of size $n =
n_1 + n_2 = 64$ is taken, without replacement, from the population $\u
P$. That is, $\u s = (s_1,\ldots, s_n)$, where $s_j \in\u P$
and $s_j \neq s_{j'}$. One of the two treatments will be assigned to
each of the units in the sample $\u s$. For the example, treatment~1
was assigned to $n_1=32$ and treatment~2 was assigned to $n_2 = 32$ of
the 64 sampled units.

Let $y_{t.s_j}$ be the response value for sampled subject $s_j$ when
exposed to treatment $t$. That is, $y_{t.s_j}$ is a realization of
$Y_{t.s_j}$. Of course,
for each subject $s_j$, only one of the realizations, $y_{1.s_j}$ or
$y_{2.s_j}$, will be observed.
From a potential variables viewpoint, the original data in Table~\ref{tab1} can
be viewed as follows:

\begin{remark*}
Table~\ref{tab1} used the conventional $y_{t,i}, i=1,\ldots, 32,
t=1,2$ to represent the observed data, whereas Table~\ref{tab2} uses $y_{2.s_1},
y_{2.s_2}, y_{1.s_3}, \ldots, y_{2.s_{63}}, y_{1.s_{64}}$. It is
important to note that the symbols $y_{t,i}$ and $y_{t.i}$ represent
very different objects.
For example, in Table~\ref{tab1}, of those units sampled and assigned treatment
1, the 3rd had a process value of $y_{1,3}$, and of those
units sampled and assigned treatment 2, the 3rd had a process value of
$y_{2,3}$. That is, $y_{1,3}$ and $y_{2,3}$ are process
values for two \emph{distinct units}. In contrast, $y_{1.3}$ and
$y_{2.3}$ represent the process values under treatments 1 and 2 for the
\emph{same unit}, specifically, the 3rd unit in the population.
\end{remark*}

\section{Data-Generation Models and Inference Goals}\label{sec3}

Let
$\u Y = (Y_{1.1},\ldots, Y_{1.N}, Y_{2.1}, \ldots, Y_{2.N})$ be the
vector of potential variables for the population $\u P$ and
$\u y = (y_{1.1}, y_{1.2},\ldots, y_{1.N}, y_{2.1},\ldots, y_{2.N})$ be
the corresponding vector of realizations. We will use this notational
convention throughout the paper: upper case letters for random
variables and lower case letters for realizations.

To simplify and to highlight vector component identification, we introduce
dot ``.'' operations and a component-selection bracket ``[ ]'' notation
that is similar to the matrix syntax used in computer languages such
as R.
Let $\u x$ and $\u w$ be $m$-dimensional vectors and let $k$ be a scalar.
Define
\begin{eqnarray*}
\u x.\u w &=& (x_1.w_1,\ldots,x_m.w_m)\quad
\mbox{and}\\
k.\u x &=& (k.x_1,\ldots,k.x_m).
\end{eqnarray*}
Consider an $m$-dimensional vector $\u x$ with components identified by
subscripts $a_1,\ldots, a_m$, that is, $\u x = (x_{a_1},\ldots,
x_{a_m})$. Provided $\u b = (b_1,\ldots, b_q)$ has components $b_i \in
\{a_1,\ldots, a_m\}$, for each $i=1,\ldots, q$, the vector $\u x[\u b]$
is defined as $\u x[\u b] = \u x[b_1,\ldots, b_q] =(x_{b_1},\ldots,  x_{b_q})$.

As an example, $\u y = (y_{1.1},\ldots, y_{1.N}, y_{2.1},\ldots,
y_{2.N})$ can be expressed as $\u y = \u y[1.\u P, 2.\u P]$. Similarly,
$\u y[1.\u s] = (y_{1.s_1},\ldots, y_{1.s_n})$ and $\u y[\u t.\u
s] = (y_{t_1.s_1},\ldots,  y_{t_n.s_n})$.
We will also use a notation for averages: As examples,
\begin{eqnarray*}
\overline Y[t.\u P] &=& N^{-1}\sum_{i=1}^N
\u Y[t.i], \\
\overline y[t.\u P] &=& N^{-1}\sum
_{i=1}^N\u y[t.i]\quad \mbox{and}\\
\overline y[t.\u s]& =&
n^{-1}\sum_{j=1}^n \u
y[t.s_j].
\end{eqnarray*}

The data-generation models we consider in this paper are based on the
following sequential generations:
%
\begin{eqnarray}\label{dgm}
\u y \leftarrow \u Y \nonumber\\
\eqntext{\mbox{here, } \u y =
(y_{1.1},\ldots, y_{1.N}, y_{2.1},\ldots,
y_{2.N}),}
\\
\u s \leftarrow\u S |(\u Y = \u y)
\nonumber
\\[-8pt]
\\[-8pt]
\eqntext{\mbox{here, } \u s = (s_1,
\ldots, s_n), s_j \in\u P, s_j \neq
s_{j'},}
\\
\u t \leftarrow \u T |(\u Y = \u y, \u S = \u s)\nonumber \\
\eqntext{\mbox {here, } \u t =
(t_1,\ldots, t_n), t_j \in\{1,2\}.}
\end{eqnarray}
The left arrow ``$\leftarrow$'' is read, ``is a realization of.'' Here,
$\u Y$ is the collection of $2N$ potential response variables, $\u S$
is the collection of $n$ sampling variables, and $\u T$ is the
collection of $n$ treatment assignment variables.
The sequencing in (\ref{dgm}) is not required to correspond to the
temporal sequencing of data generation. It is meant only as a device
for specifying the joint distribution of $(\u Y, \u S, \u T)$. For a
related discussion, see Rubin [\citeyear{RUB10}, between equations (4) and (5)].

In words, ``Nature'' generates $N$ units, which are labeled $1,2,\ldots,
N$. Each unit can potentially experience either of two ``possible
worlds,'' which correspond to exposure under the two treatments.
The vector $\u y$ contains the $2N$ potential response values, one for
each of the $N$ units under treatment 1 and one for each of the $N$
units under treatment 2. These potential deviates in $\u y$ are viewed
as realized, at least in theory, but only partially observable. We
sample $n$ distinct subjects $\u s$ from the population $\u P$. The
sampling may depend on potential deviates $\u y$; this dependence often
stems from selecting on covariates that are statistically related to
the potential variables [see \citett{RUB10}, between equations (4) and
(5)]. Finally, we assign treatment levels $\u t$ to units in the
sample, that is, we choose which of the two possible worlds we will
observe for each unit in the sample. The treatment assignment may
depend on the potential deviates $\u y$ and/or the sampled units $\u
s$. However, when mechanical or physical randomization (cf. \citett{FIS35,Kem55}) is used, the treatment assignment can be made
to be independent of the potential deviates.

We will refer to the potential variables $\u Y$ as ``process
variables''\footnote{\citet{Rub05}, uses the descriptor ``science'' rather
than ``process.''} and the values $\u y$ as ``process values,'' to
differentiate them from the ``selection'' variables $(\u S, \u T)$ and
values $(\u s, \u t)$.
The process portion describes how things behave under both possible
worlds and the selection portion determines how we go about observing
this behavior. Owing to the sampling and treatment assignment (the
selection), we do not observe the entire vector of potential deviates
$\u y$ (the process values). Indeed, the ``fundamental problem of
causal inference'' rules out the possibility of fully observing the
$2N$-dimensional data vector $\u y$. Instead, we observe only the
$n$-dimensional sub-vector
$
\u y[\u t.\u s]. $
Schematically, we have
\begin{eqnarray*}
\underbrace{\underline y[\underline t.\underline s]}_{\mathrm{ observed}}
\subseteq
\underbrace{\underline y[1.\underline s, 2.\underline s] \subseteq \underline y[1.\underline
P, 2.\underline P] = \underline y \leftarrow \underline Y}_{\mathrm {unobserved}}.
\end{eqnarray*}
The inference goal of this paper can be stated succinctly as follows:

\textit{Inference goal}. Use the observed data $\u y[\u t.\u
s]$ from a comparative experiment to reduce uncertainty about one of
the three targets: the vector $\u y[1.\u s, 2.\u s]$, the vector
$\u y[1.\u P, 2.\u P]$, or the distribution of $\u Y$.

\section{Treatment Effects and ``No-Treatment-Effect'' Hypotheses}\label{sec4}

\subsection{Treatment Effects}\label{sec4.1}

We began this paper with the question of whether there was a treatment
effect. Of course, this raises another question: What exactly is a
``treatment effect''?

In a comparative experiment, a treatment effect can be viewed as some
measure of the difference between the response ($\u Y$) distribution or
response values ($\u y$) for treatment level 1
and the response distribution or response values for treatment level 2.
The potential variables viewpoint lends itself to intuitively-appealing
candidate definitions of such treatment effects (cf. Neyman, \citeyear{Ney90};
Rubin, \citeyear{Rub90}, \citeyear{Rub05}, \citeyear{RUB10}).
Some of the candidates considered in this paper are as follows:

\begin{tabular}{l}
Realized unit-specific effects:\\
\quad$ \u y[1.s_j]
- \u y[2.s_j], j=1,\ldots,n$ or\\
\quad$\u y[1.i] - \u y[2.i],
i=1,\ldots,N$.\\
Distribution unit-specific effects:\\
\quad$\delta(F_{1.i}, F_{2.i}), i=1,\ldots, N$.
\\
Expected unit-specific effects: \\
\quad$E(\u Y[1.i]) - E(\u
Y[2.i]), i=1,\ldots, N$.\\
Realized aggregate effects:\\
\quad$\overline y[1.\u s]
- \overline y[2.\u s]$ or\\
\quad$\overline y[1.\u P] -
\overline y[2.\u P]$.\\
Expected aggregate effects:\\
\quad$E(\overline Y[1.\u P])
- E(\overline Y[2.\u P])$.\\
\end{tabular}\vspace*{6pt}

For example, the realized unit-specific treatment effect $\u y[1.s_j] -
\u y[2.s_j]$ is simply the difference between unit $s_j$'s responses
under two scenarios or two possible worlds---in one world the unit
is exposed to treatment~1 and in the other world the unit is exposed to
treatment~2. As another example, the distribution unit-specific effect
$\delta(F_{1.i}, F_{2.i})$ measures the distance between the c.d.f.'s of
$\u Y[1.i]$ and $\u Y[2.i]$ using some distance function $\delta(\cdot
)$. This latter example illustrates that treatment effects need not be
defined in terms of simple differences, arithmetic averages, or means
of distributions. Other examples of treatment effects include the
distribution unit-specific effect $\operatorname{median}(\u Y[1.i]) - \operatorname{median}(\u
Y[2.i])$, realized unit-specific effects, such as $(\u y[2.s_j] - \u
y[1.s_j])/\u y[1.s_j]$, and realized aggregate effects, such as $\|
\u y[1.\u s] - \u y[2.\u s]\|$ or $\operatorname{var}(\u y[1.\u s]) -
\operatorname{var}(\u y[2.\u s])$ or ${\overline y[2.\u s] -
\overline y[1.\u s] \over\overline y[1.\u s]}$, or
for binary responses, the realized odds ratio $
{\overline y[1.\u s]/(1- \overline y[1.\u s])\over\overline y[2.\u
s]/(1-\overline y[2.\u s])}$, etc.\vspace*{1pt}

Unfortunately, none of the treatment effects mentioned above is
observable. The expected and distribution effects cannot be observed
because the distribution of $\u Y$ is not completely known. The
realized effects cannot be observed because, by the fundamental problem
of causal inference, only one of the realizations, for example, either
$\u y[1.s_j]$ or $\u y[2.s_j]$, can be observed. Fortunately, this does
not preclude unbiased estimation of some of these unobservable
treatment effects, as we point out below.

In the potential-variables causal literature, the treatment effects
defined above would be considered causal effects provided certain
assumptions hold (e.g.,
Rubin, \citeyear{Rub90}, \citeyear{Rub05}, \citeyear{RUB10}). To avoid the ongoing debate about the nature
of causality, we will refrain from referring to treatment effects as
causal effects.

\subsection{``No-Treatment-Effect'' Hypotheses}\label{sec4.2}

Corresponding to each treatment effect definition, there is a
``no-treatment-effect'' hypothesis. As examples,
\begin{eqnarray}
&&H_0^{UP}: \u Y[1.\u P] = \u Y[2.\u P],\quad \mbox{with probability 1};\nonumber\\
&&\hspace*{32pt}H_{0}^{DUP}: \u Y[1.i] \sim\u Y[2.i],
i=1,\ldots, N.\nonumber\\
\eqntext{\mbox{Herein, ``$\sim$'' means ``distributed
as'';}}\\
&&\hspace*{70pt}H_{0}^{EUP}: E(\u Y[1.i]) = E(\u Y[2.i]),\nonumber\\
\eqntext{ i=1,\ldots
, N;}\\
&&\hspace*{32pt}H_{0}^{RUP}: \u y[1.\u P] = \u y[2.\u P];\nonumber\\
&&\hspace*{70pt}H_{0}^{RAP}: \overline y[1.\u P] = \overline y[2.\u
P];\nonumber\\
&&\hspace*{70pt}H_{0}^{RUs}: \u y[1.\u s] = \u y[2.\u s];\nonumber\\
&&\hspace*{105pt}H_{0}^{RAs}: \overline y[1.\u s] = \overline y[2.\u
s].\nonumber
\end{eqnarray}

The indentations are used to denote nesting. For example, both
$H_0^{EUP}$ and $H_0^{RUP}$ are implied by $H_0^{UP}$.
Similarly, $H_0^{RAs}$ is implied by $H_0^{RUs}$ and by $H_0^{RUP}$,
but not by $H_0^{RAP}$. The superscripts remind us of the type of
treatment effect used in the hypothesis. For example, the hypothesis
$H_0^{EUP}$ uses \emph{E}xpected \emph{U}nit-specific effects (for the
\emph{P}opulation), and $H_0^{RAs}$ uses \emph{R}ealized \emph
{A}ggregate (over \emph{s}ample $\u s$) effects.

\section{Inference Approaches and Assumptions}\label{sec5}
The $(\u y, \u s, \u t)$ components in the observed data $\u y[\u t.\u
s]$ are viewed as outcomes of the sequential generations of (\ref
{dgm}). The complete, but only partially observed, data $\u y$ is a
realization of the $2N$-dimensional vector of potential variables $\u
Y$. In symbols, we have
$\u y[\u t.\u s] \leftarrow \u Y[\u T.\u S]$ and $\u y \leftarrow\u Y$.

As stated previously, the inference goal is to use the observed data
$\u y[\u t.\u s]$ to reduce uncertainty about one of three targets: the
distribution of $\u Y$, the vector $ \u y[1.\u P, 2.\u P]$, or the
vector $\u y[1.\u s, 2.\u s]$.
The choice of inference approach depends on which of these targets we
are interested in and it depends on what assumptions we can reasonably
make about the joint distribution of $(\u Y, \u S, \u T)$, where $\u Y$
is the ``process'' variable and $(\u S, \u T)$ are the ``selection''
variables. More specifically, $\u S$ is the ``sampling'' variable and
$\u T$ is the ``randomization'' or treatment assignment variable. In
this paper, we consider three candidate inference approaches.

\subsection{Process-Based Inference and Assumptions}\label{sec5.1}
With the process-based approach, we condition on the selection (only
$\u Y$ is random) and use
\begin{eqnarray*}
\u y[\u t.\u s] &\leftarrow &\u Y[\u T.\u S] | (\u S = \u s, \u T = \u t)\\
& \sim& \u
Y[\u t.\u s] | (\u S = \u s, \u T = \u t)
\end{eqnarray*}
to carry out inferences about the distribution of $\u Y$. (The
discussion section describes more general inferences.)

With process-based inference, we generally must make assumptions about
the conditional distribution of $\u Y|(\u S = \u s, \u T = \u t)$.
However, because this paper focuses on test procedures that are valid
provided the process is independent of the selection (see assumption
$A_1$), we will only make assumptions about the (unconditional) process
distribution of $\u Y$; see assumptions $A_2$--$A_7$.
\begin{eqnarray*}
&&A_1: (\u S, \u T)  \perp\!\!\!\perp\u Y;
\\
&&A_2: \u Y[t.i] \sim F_t,\quad i=1,\ldots, N, t=1,2;
\\
&&A_3: \u Y[1.i,2.i],\quad  i=1,\ldots, N,\mbox{ are independent};
\\
&&A_4: F_t  \in \{\mbox{continuous c.d.f.s}\}; \\
&&A_5: F_t  \in  \{N(\mu_t, \sigma_t^2)  \mbox{ c.d.f.s}\};\\
&& A_6: F_t  \in \{N(\mu_t, \sigma^2)  \mbox{ c.d.f.s}\};
\\
&&A_7: F_t  \in \{\mbox{c.d.f.s with mean and variance }  (\mu_t, \sigma
^2_t)\}.
\end{eqnarray*}

Process-based inference is simplified under Assumption $A_1$ because
the observed data can be viewed as a realization of $\u Y[\u t.\u s]$;
in symbols, $\u y[\u t.\u s] \leftarrow\u Y[\u t.\u s]$. It follows
that we need only model the (unconditional) distribution of the process
variable $\u Y$. Importantly, under $A_1$, process-based inference does
not require any assumptions about the selection $(\u S, \u T)$. It is
also important to note that when $A_1$ holds and $\u Y$ is modeled via
assumptions such as $A_2$--$A_7$, the observed data $\u y[\u t.\u s]$
can be used to make process-based inferences about the distribution of
$\u Y$.

Unfortunately, Assumption $A_1$ is typically not tenable in practice.
Notice that $A_1$ is equivalent to the two assumptions, $\u T \perp\!\!
\!\perp\u Y|\u S$ and $\u S \perp\!\!\!\perp\u Y$.
When mechanical randomization is used to assign treatments to the
sampled units, the first assumption can be made tenable.
However, the reasonableness of the second assumption, that the sampling
variable $\u S$ is independent of $\u Y$, is often questionable in
practice. For example, with haphazard or convenience sampling, rather
than probability sampling, it often turns out that $\u S$ and $\u Y$
are not independent. The dependence typically stems from sampling on
the basis of covariates that are related to $\u Y$.\footnote{Of course,
if the covariates responsible for the dependence were known and
observable, we could condition on their values to restore independence;
however, this conditional model falls outside the purview of the
current paper.}

The assumption $A_2$ is not as restrictive as it may initially appear.
For example, whenever the identifiers are arbitrarily assigned to the
$N$ population units, the $N$ pairs $\u Y[1.i, 2.i]$ would be
exchangeable and, hence, $A_2$ would hold. Generally, the more tenuous
assumptions are $A_1$, that the selection is carried out independently
of the process, the independence assumption $A_3$, and the assumptions
$A_4$--$A_7$ about the form of marginal distributions $F_1$ and $F_2$.

\subsection{Randomization-Based Inference and Assumptions}\label{sec5.2}
With the randomization-based approach, we condition on both the process
values and the sample (only $\u T$ is random) and use
\begin{eqnarray*}
\u y[\u t.\u s] &\leftarrow& \u Y[\u T.\u S] | (\u Y = \u y, \u S = \u s) \\
&\sim& \u
y[\u T.\u s] | (\u Y = \u y, \u S = \u s)
\end{eqnarray*}
to carry out inferences about $\u y[1.\u s, 2.\u s]$.

With randomization-based inference, we generally must make assumptions
about the conditional distribution of $\u T|(\u Y = \u y, \u S = \u
s)$. However, because this paper focuses on test procedures that are
valid provided the randomization is conditionally independent of the
process, given the sample (see assumption $B_1$), we will only make
assumptions about the distribution of $\u T|(\u S = \u s)$; see
assumption~$B_2$.
\begin{eqnarray}
&&B_1:  \u T  \perp\!\!\!\perp \u Y  |  \u S.\nonumber\\
&& B_2: \mbox{ The distribution of $\u T|(\u S = \u s)$ is completely}\nonumber\\
&&\hspace*{24pt}\mbox{known and satisfies\ldots}\nonumber\\
&&\hspace*{24pt}\quad P(\u T.\u S \ni t.s_j  |  \u S = \u s)  >  0,\nonumber\\
 \eqntext{j=1,\ldots, n,
  t=1,2;}\\
&&\hspace*{24pt}\quad P(\u T.\u S \ni t.s_j,  \u T.\u S \ni t'.s_{j'}  |  \u S
= \u s)   =  0     \nonumber\\
 \eqntext{\mbox{if and only if } t \neq t'
 \mbox{and}   j = j'.}
 \end{eqnarray}

Randomization-based inference is simplified under
assumption $B_1$ and fortunately the use of mechanical randomization
makes this assumption tenable. Under $B_1$, we have that the observed
data can be viewed as $\u y[\u t.\u s] \leftarrow \u y[\u T.\u s]|(\u
S = \u s)$, so only the distribution
of $\u T|(\u S = \u s)$ needs to be modeled.
In particular, we need not make any assumption about the distribution
of $\u Y$ or its relation to $\u S$; for example, $\u Y$ and $\u S$,
that is, the process and the sampling, need not be independent.
It is important to note that when the distribution of $\u T|(\u S = \u
s)$ is completely known (see $B_2$), the distribution of $\u y[\u T.\u
s]|(\u S = \u s)$ is known up to the partially-observed values $\u y
[1.\u s, 2.\u s]$, which are the parameters of interest for
randomization-based inference. Thus, when $B_1$ and $B_2$ hold, the
observed data $\u y[\u t.\u s]$ can be used to carry out
randomization-based inference about the target parameters $\u y[1.\u s,
2.\u s]$.

In $B_2$, the probabilities are called first- and second-order
inclusion probabilities for the random sample, namely, $\u T.\u S|(\u S
= \u s)$, taken from $(1.\u s, 2.\u s)$. Assumption $B_2$ imposes
constraints on these inclusion probabilities. The positive first-order
inclusion probabilities imply that ``proper'' randomization is used to
assign treatments, that is, each unit in the sample has a positive
probability of receiving either treatment; we say that this is a ``\emph
{proper}'' \emph{randomized comparative experiment}.\footnote{The
two-treatment completely randomized design (CRD) experiment is a
special-case example of a proper randomized comparative experiment.
With the CRD, $\u T|(\u S=\u s)$ has
a uniform distribution over all possible rearrangments of $n_1$ $1$'s
and $n_2$ $2$'s (cf. Cox, \citeyear{Cox92}, pages~71--72;
\citett{Kem77}, Section~\ref{sec8}; or \citett{COXREI00}, Section~2.2.4.)} Put another way, $\u T.\u S|(\u S = \u s)$
is a \emph{{probability}} sample from $(1.\u s, 2.\u s)$. Because the
same unit cannot be assigned different treatments, the second-order
inclusion probabilities with $t \neq t'$ and $j = j'$ are 0. This
implies that the
probability sample is nonmeasurable, to use language from sampling
theory (cf. S\"arndal et al., \citeyear{SarSweWre92}, pages~32--33). This nonmeasurability
complicates the
computation of certain randomization-based test statistics (see
Section~\ref{sec6.3.2} below), as Neyman was fully aware of in 1923.

\subsection{Selection-Based Inference and Assumptions}\label{sec5.3}
With the selection-based approach, we condition on the process values
[only $(\u S, \u T)$ is random] and use
\[
\u y[\u t.\u s] \leftarrow \u Y[\u T.\u S] | (\u Y = \u y) \sim \u y[\u T.\u S]
| (\u Y = \u y)
\]
to carry out inferences about $ \u y[1.\u P, 2.\u P]$.

With selection-based inference, we generally must make assumptions
about the conditional distribution of $(\u S, \u T)|(\u Y = \u y)$.
However, because this paper focuses on test procedures that are valid
provided the selection is independent of the process (see assumption
$C_1$), we will only make assumptions about the unconditional
distribution of $(\u S, \u T)$; see assumption $C_2$.
\begin{eqnarray}
&&C_1:  (\u S, \u T)  \perp\!\!\!\perp \u Y.\nonumber\\
&&C_2: \mbox{The distribution of $(\u S, \u T)$ is completely known}\nonumber\\
&&\hspace*{21pt}\mbox{and satisfies\ldots}\nonumber\\
&&\hspace*{21pt}\quad P(\u T.\u S \ni t.i )  >  0,\nonumber\\
 \eqntext{i=1,\ldots, N,   t=1,2.}\\
&&\hspace*{21pt}\quad P(\u T.\u S \ni t.i,  \u T.\u S \ni t'.i' )   =  0\nonumber  \\
\eqntext{\mbox{if and only if } t \neq t'  \mbox{ and }   i = i'.}
\end{eqnarray}

Selection-based inference is simplified under assumption $C_1$ because
the observed data can be viewed as $\u y[\u t.\u s] \leftarrow\u y[\u
T.\u S]$. It follows that we need only specify the (unconditional)
distribution of the selection $(\u S, \u T)$; no assumptions about $\u
Y$ are needed. Unfortunately, as discussed in the process-based
subsection above, assumption $C_1$ is not usually tenable in practice
because the sampling and process are often dependent. It is important
to note that when the distribution of $(\u S, \u T)$ is completely
known (see $C_2$), the distribution of $\u y[\u T.\u S]$ is known up to
the partially-observed values $\u y [1.\u P, 2.\u P]$, which are the
parameters of interest for selection-based inference. Thus, when $C_1$
and $C_2$ hold, the observed data $\u y[\u t.\u s]$ can be used to
carry out selection-based inference about the target parameters $\u
y[1.\u P, 2.\u P]$.

As discussed in the randomization-based section, assumption $C_2$
imposes constraints on first- and second-order inclusion probabilities.
In this case, the random sample $\u T.\u S$ is taken from $(1.\u P,
2.\u P)$. The assumption implies
that each of the $2N$ elements in $(1.\u P, 2.\u P)$ has a positive
probability of being selected. Thus, the random sample is a probability
sample. The 0 second-order inclusion probabilities imply that the
probability sample is nonmeasurable.

\section{Tests of the No-Treatment-Effect Hypothesis}\label{sec6}

This section describes a collection of process-, randomization-, and
selection-based tests of no treatment effect hypotheses. Some of these
tests are well known (e.g., the two-sample $t$ test), and some are less
well known (e.g., the Neyman randomization test). In any case, we will
emphasize the assumptions needed for their applicability and we will
carefully state the hypothesis that is actually being tested. We begin
by introducing a difference statistic that forms the basis of most of
the tests considered in this paper.

\subsection{The Difference Statistic}\label{sec6.1}

With the exception of the Wilcoxon rank sum statistic, this paper will
focus on test statistics that are based
on the following difference statistics:
\begin{eqnarray*}
&&\underbrace{D_1\bigl(\u Y[\u t.\u s]\bigr) = D({
\u Y}, \u s, \u t, \u w_1)}_{\mathrm{ process}}, \\
&&\underbrace{D_{3}(\u T) = D(\u y, \u s,{
\u T}, \u w_3)}_{\mathrm{randomization}}, \\
&&D\underbrace{_{23}(\u S,\u T) = D(\u y, {\u S, \u T},
\u w_{23})}_{\mathrm{selection}},
\end{eqnarray*}
where
%
\begin{eqnarray}\label{D}
D(\u y, \u s, \u t, \u w) &=& \underbrace{\sum_{i=1}^N
{\u y[1.i] \i(\u t.\u s \ni1.i)\over\u w[1.i]}}_{\mathrm{
weighted\ avg\ of\   trt\  1\  values}}
\nonumber
\\[-8pt]
\\[-8pt]
\nonumber
&&{}- \underbrace{\sum
_{i=1}^N {\u
y[2.i] \i(\u t.\u s \ni2.i)\over\u w[2.i]}}_{\mathrm{weighted\ avg\ of\ trt\
2\ values}}.
\end{eqnarray}
The candidate values for weights $\u w$ include
\begin{eqnarray*}
\u w_1[t.i] &=& n_t,\qquad \u w_{3}[t.i] = nP(\u T.
\u s \ni t.i | \u S= \u s),\\
 \u w_{23}[t.i] &=& NP(\u T.\u S \ni t.i),
\end{eqnarray*}
where $n = \mathrm{length}(\u s),  n_t = {\sum_{j=1}^n \i(t_j =
t)}$. From the discussions in Sections~\ref{sec5.2} and \ref{sec5.3}, it follows that the
$\u w_{3}$ and $\u w_{23}$ components are multiples of first-order
inclusion probabilities (cf. S\"arndal et al., \citeyear{SarSweWre92}), using language
from finite-population sampling theory. By convention, we set $0/0
\equiv0$ in (\ref{D}).

There are several useful properties of these $D$ statistics.
First, note that
\[
\begin{tabular}{p{200pt}@{}}
$D(\u y,\u s,\u t, \u w)$ can be computed using only the observed values
$\u
y[\u t.\u s]$, $\u s$, and $\u t$.
\end{tabular}
\]
That is, $D$, and hence each of $D_1, D_3$, and $D_{23}$, is an
observable statistic. It also follows that the process-based
statistic $D_1$ depends on $\u Y$ only through $\u Y[\u t.\u s]$, hence
the notation $D_1(\u Y[\u t.\u s])$.
Second, the process-based statistic $D_1$ is simply the difference
between the unweighted sample averages $n_1^{-1}\sum_{j: t_j = 1} \u
Y[1.s_j]$ and $n^{-1}_2\sum_{j:t_j=2} \u Y[2.s_j]$. The randomization-
and the selection-based statistics $D_3$ and $D_{23}$ are differences
between probability-weighted sample averages.
Third,
%
\begin{eqnarray}\label{Ddistn}
&&\mbox{Under $A_1$,} D_1|(\u S = \u s, \u T = \u t)\nonumber\\
&&\quad \mbox{has distribution that depends only on the model}\nonumber\\
&&\quad\mbox{for }\u Y[\u t.\u s].\nonumber\\
&&\mbox{Under $B_1$,} D_3|(\u Y = \u y, \u S = \u s)\nonumber\\
&&\quad\mbox{has distribution that depends only on the}\\
&&\quad\u y[1.\u s, 2.\u s] \mbox{ values}
\mbox{ and the $\u T|(\u S = \u s)$ distribution.}\nonumber\\
&&\mbox{Under $C_1$,} D_{23}|(\u Y = \u y)\nonumber\\
&&\quad \mbox{has distribution that depends only on the}\nonumber\\
&&\quad \u y[1.\u P, 2.\u P] \mbox{ values}
\mbox{ and the $(\u S, \u T)$ distribution.}\nonumber
\end{eqnarray}
Fourth,
%
\begin{eqnarray}\label{ED}
&&\mbox{Under $A_1$ and
$A_2$,} \nonumber\\
&&\quad E(D_1|\u S = \u s, \u T = \u t) =
E(D_1) = \mu_1 - \mu_2.\nonumber
\\
&&\mbox{Under $B_1$ and $B_2$,}\nonumber\\
&&\quad E(D_3|\u Y
= \u y, \u S = \u s)  =  E(D_3| \u S = \u s)\\
&&\hspace*{104pt} = \overline y[1.\u
s] - \overline y[2.\u s].\nonumber
\\
&&\mbox{Under $C_1$ and $C_2$,}\nonumber\\
&&\quad E(D_{23}|\u Y
= \u y) = E(D_{23}) = \overline y[1.\u P] - \overline y[2.\u P].\nonumber
\end{eqnarray}
Here, $\mu_t = E(\u Y[t.i])$ is the mean of the assumed common
distribution $F_t$. The last two expectation results
follow because $D_3$ and $D_{23}$ are Horvitz--Thompson
probability-weighted estimators\break (\citett{HorTho52}; S\"arndal
et al., \citeyear{SarSweWre92}, page~43). These expectation results highlight the usefulness of
basing tests of ``no treatment effects'' on these $D$ statistics, at
least when the treatment effect is measured in terms of differences in
means. These results also highlight the usefulness of random sampling
and treatment randomization.

\subsection{Process-Based Tests}\label{sec6.2}

With the process-based approach, we condition on the selection (only
$\u Y$ is random) and use
\begin{eqnarray*}
\u y[\u t.\u s]& \leftarrow& \u Y[\u T.\u S] | (\u S = \u s, \u T = \u t) \\
&\sim& \u
Y[\u t.\u s] | (\u S = \u s, \u T = \u t)
\end{eqnarray*}
to carry out inferences about the distribution of $\u Y$. Among other
assumptions, the validity of the process-based tests described below
generally require that assumptions $A_1$: $(\u S, \u T) \perp\!\!\!
\perp\u Y$;
$A_2$: $\u Y[t.i]  \sim F_t, i=1,\ldots,N, t=1,2$; and $A_3$: $\u
Y[1.i, 2.i], i=1,\ldots,\break N$ \emph{are independent} hold. As noted in
Section~\ref{sec5.1}, these assumptions are often untenable in practice,\footnote
{The tests are often invalid
because the selection is related to the process, the treatment-specific
process variables are not identically distributed, and/or the process
variables are not
independent across units.}
so the reader is reminded to
apply these tests with caution.

\subsubsection{Permutation test}\label{sec6.2.1}

Consider the no-treatment-effect hypothesis
\[
H_0^{DUP}: \u Y[1.i] \sim \u Y[2.i],\quad i=1,\ldots,N.
\]
Under $H_0 = (A_1, A_2, A_3, H_0^{DUP})$, we can state the null as
$H_0^{DUP}$: $F_1 = F_2$ and base our test on
\begin{eqnarray}
D_1| \bigl(\u Y[\u t.\u s] \in\Pi\bigl(\u y[\u t.\u s]\bigr)\bigr)\nonumber\\
\eqntext{\mbox{which has a known,}}\\
\eqntext{\mbox{computable distribution under $H_0$}.}
\end{eqnarray}
Here $\Pi(\u x) = \{\mbox{set of distinct permutations of $\u x$}\}$.\break 
That this distribution is known under $H_0$ follows because
in this case
%
\begin{eqnarray}\label{permy}
&&\u Y[\u t.\u s] | \bigl(\u Y[\u t.\u s] \in\Pi\bigl(\u y[\u t.\u s]\bigr)\bigr)
\nonumber
\\[-8pt]
\\[-8pt]
\nonumber
&&\quad\stackrel{H_0} {\sim} \mbox{uniform over points in $\Pi\bigl(\u y[\u
t.\u s]\bigr)$}.
\end{eqnarray}
The computability follows because $D_1(\u x)$ can be computed for any
$\u x \in\Pi(\u y[\u t.\u s])$.

In practice, we would report a one- or two-sided $p$-value. For example,
letting $D_{1,\mathrm{obs}} = D_1(\u y[\u t.\u s])$ be the observed difference,
a two-sided
$p$-value can be defined as
\begin{eqnarray*}
&&\operatorname{pval}(D_{1,\mathrm{obs}}) \\
&&\quad= P_{H_0}\bigl(|D_1| \geq
|D_{1,\mathrm{obs}}| | \u Y[\u t.\u s] \in\Pi\bigl(\u y[\u t.\u s]\bigr)\bigr).
\end{eqnarray*}
The size of the test that rejects $H_0$ iff $\operatorname{pval} \leq\alpha$ is less
than or equal to $\alpha$. If we observe a $p$-value $\leq\alpha$ and we
assume that $A_1, A_2$, and $A_3$ hold, then we have statistical
evidence against $H_0^{DUP}$: $F_1 = F_2$, that is, evidence at the
$\alpha$ level that $F_1 \neq F_2$.

Remark: At first glance, one might think that exchangeability of the
$N$ pairs $\u Y[1.i, 2.i]$ could replace $(A_2, A_3)$. Unfortunately, a stronger
exchangeability assumption would be needed to guarantee the uniform
permutation distribution of (\ref{permy}). Specifically, the assumption
must lead to the exchangeability of the $n$ components of $\u Y[\u t.\u
s]$. Along these lines, we
could consider a more restrictive no-treatment-effect hypothesis, for
example, $H_0^{DUP*}$: all $2N$ components in $\u Y[1.\u P, 2.\u P]$
are exchangeable.
Then the permutation test would be valid under $H_0^* = (A_1, H_0^{DUP*})$.
It is useful to note that $H_0^{DUP*}$ can be viewed as $H_0^{DUP}$
along with extra assumptions about the process distribution. In this
sense, this strong exchangeability hypothesis is an example of the
no-treatment-effect hypotheses considered herein.

\subsubsection{Wilcoxon rank sum test}\label{sec6.2.2}

Consider the no-treatment-effect hypothesis
\[
H_0^{DUP}: \u Y[1.i] \sim \u Y[2.i],\quad i=1,\ldots,N.
\]
Under $H_0 = (A_1, A_2, A_3, A_4, H_0^{DUP})$, where $A_4$ is the
assumption that the c.d.f.s $F_t$ are continuous, we can state the null as
$H_0^{DUP}$: $F_1 = F_2$ and base our test on
\begin{eqnarray}
W \equiv W(\u R) = \sum_{j=1}^n
R_j \i(t_j = 1),\nonumber\\
\eqntext{\mbox{where } W| \bigl(\u R \in\Pi(\u
r)\bigr)}\\
\eqntext{\mbox{has a known, computable distribution under $H_0$}.}
\end{eqnarray}
Here $R_j = \operatorname{rank}(\u Y[t_j.s_j])$ and $r_j = \operatorname{rank}(\u y[t_j.s_j])$, where
the ranks are taken over the $n$ values in $\u Y[\u t.\u s]$ and
$\u y[\u t.\u s]$, respectively. Again, the $\Pi(\u r)$ is the set of
permutations of $\u r$. That this distribution is known under $H_0$
follows because
in this case
%
\begin{equation}\quad
\u R | \bigl(\u R \in\Pi(\u r)\bigr) \stackrel{H_0} {\sim} \mbox
{uniform over points in $\Pi(\u r)$}. \label{permr}
\end{equation}
The computability follows because $\u R(\u x)$ can be computed for any
$\u x \in\Pi(\u r)$.

In practice, we would report a one- or two-sided $p$-value. For
convenience, let $W_{\mathrm{obs}} = W(\u r)$ be the observed rank sum
statistic. Then the following is a reasonable two-sided $p$-value (of
course there are others):
\begin{eqnarray*}
\operatorname{pval}(W_{\mathrm{obs}}) &= &2\min\bigl\{P_{H_0}\bigl(W \geq
W_{\mathrm{obs}}| \u R \in\Pi(\u r)\bigr),\\
&&\hspace*{28pt} P_{H_0}\bigl(W \leq
W_{\mathrm{obs}}| \u R \in\Pi(\u r)\bigr)\bigr\}.
\end{eqnarray*}
Assuming that $A_1$--$A_4$ hold, an observed $p$-value $\leq\alpha$
would give us
statistical evidence against $H_0^{DUP}$: $F_1 = F_2$, that is, evidence
at the $\alpha$ level that $F_1 \neq F_2$.

\subsubsection{Two-sample $t$ tests}\label{sec6.2.3}

Consider the no-\break treatment-effect hypothesis
\[
H_0^{EUP}: E\bigl(\u Y[1.i]\bigr) \sim E\bigl(\u Y[2.i]
\bigr),\quad i=1,\ldots,N.
\]
Under $H_0 = (A_1, A_2, A_3, A_5, H_0^{EUP})$, where $A_5$ states that
we are sampling from $N(\mu_t, \sigma^2_t)$ distributions, we can state
the null as $H_0^{EUP}$: $\mu_1 = \mu_2$ and base our test on
\[
T \equiv{D_1\over \mathit{SE}(D_1)} \stackrel{H_0} {\sim} \operatorname{approx} t(
\nu),
\]
where $\nu$ is Welch's formula for the approximate degrees of freedom
and $t(\nu)$ is Student's $t$ distribution. The standard error has the
familiar form
\[
\mathit{SE}(D_1) = \sqrt{{\what\sigma_1^2\over n_1} +
{\what\sigma_2^2\over n_2}},
\]
where $\what\sigma_t^2$ is the sample variance of the $\{\u Y[t.s_j]:
  t_j = t\}$. Because $D_1$ is simply the difference between the two
unweighted sample averages, this statistic $T$ is identical to Welch's
(\citeyear{WEL38}) version of the two-sample $t$ statistic.

An approximate two-sided $p$-value can be computed as
\[
P_{H_0}(|T| \geq|T_{\mathrm{obs}}|) \approx P\bigl(\bigl|t(\nu)\bigr|
\geq|T_{\mathrm{obs}}|\bigr) \equiv \mathrm{apval}(T_{\mathrm{obs}}).
\]
Assuming that $A_1$--$A_3$ and $A_5$ hold, an approximate $p$-value $\leq
\alpha$ gives statistical evidence against $H_0^{EUP}$: $\mu_1 = \mu_2$,
that is, evidence at the approximate $\alpha$ level that $\mu_1 \neq
\mu_2$.

Under $H_0 = (A_1, A_2, A_3, A_6, H_0^{EUP})$, where $A_6$ states that
we are sampling from $N(\mu_t, \sigma^2)$ distributions, we can state
the null as $H_0^{EUP}$: $\mu_1 = \mu_2$ and base our test on
\[
T_p \equiv{D_1\over \mathit{SE}_p(D_1)} \stackrel{H_0} {
\sim} t(\nu), \qquad \nu= n_1 + n_2 - 2.
\]
The standard error has the familiar form
\[
\mathit{SE}_p(D_1) = \sqrt{{\what\sigma^2\over n_1} +
{\what\sigma^2\over n_2}},
\]
where $\what\sigma^2$ is the pooled estimate $((n_1 - 1)\what\sigma_1^2
+ (n_2 - 1)\what\sigma_2^2)/(n_1 + n_2 - 2)$. The statistic $T_p$ is
the standard two-sample pooled $t$ statistic.

An exact two-sided $p$-value can be computed as
\begin{eqnarray*}
P_{H_0}\bigl(|T_p| \geq|T_{p,\mathrm{obs}}|\bigr) &=& P\bigl(\bigl|t(\nu)\bigr|
\geq|T_{p,\mathrm{obs}}|\bigr) \\
&\equiv &\operatorname{pval}(T_{p,\mathrm{obs}}).
\end{eqnarray*}
Assuming that $A_1$--$A_3$ and $A_6$ hold, an exact $p$-value $\leq\alpha
$ gives statistical evidence against $H_0^{EUP}$: $\mu_1 = \mu_2$, that
is, evidence at the approximate $\alpha$ level that $\mu_1 \neq \mu_2$.

Under the less restrictive assumption, $H_0 = (A_1,\break  A_2,  A_3, A_7,
H_0^{EUP})$, where $A_7$ states that we are sampling from any
distributions $F_t$ with mean $\mu_t$ and variance $\sigma_t^2$, we can
still use $T$ to test $H_0^{EUP}$: $\mu_1 = \mu_2$, but the actual size
of the tests based on the $p$-value, which uses the $t$ approximation,
may be far from the nominal $\alpha$. In practice, the approximation
is usually reasonable when $n$ is large enough to compensate for any
asymmetry in the underlying $F_t$ distributions.

\subsection{Randomization-Based Tests}\label{sec6.3}

With the randomization-based approach, we condition on both the process
and the sample  (only $\u T$ is random), and use
\begin{eqnarray*}
\u y[\u t.\u s] &\leftarrow& \u Y[\u T.\u S] | (\u Y = \u y, \u S = \u s) \\
&\sim& \u
y[\u T.\u s] | (\u Y = \u y, \u S = \u s)
\end{eqnarray*}
to carry out inferences about $\u y[1.\u s, 2.\u s]$. The\break 
randomization-based test procedures outlined below are valid provided
the assumptions $B_1$ and $B_2$ of Section~\ref{sec5.2} hold. There it was
pointed out that these two assumptions can be made tenable when the
treatment assignment is carried out by mechanical randomization.

\subsubsection{Fisher randomization test}\label{sec6.3.1}
Although R.~A. Fisher never explicitly used potential variables, several
authors, including \citet{WEL37}, Rubin (\citeyear{Rub90}, \citeyear{Rub05}), and \citet{COX09},
have suggested that he tacitly used the no-unit-specific-effects (or
sharp null) hypothesis in this randomized comparative experiment
setting. That is, it has been suggested that to Fisher the
no-treatment-effect hypothesis had the form
$
H_{0}^{RUs}$: $\u y[1.s_j] = \u y[2.s_j],   j=1,\ldots, n$,
or, in simpler notation,
\[
H_{0}^{RUs}: \u y[1.\u s] = \u y[2.\u s].
\]
When $H_0 = (B_1, B_2, H_0^{RUs})$   holds, we have by (\ref{ED})
that $E(D_3|\u S = \u s) \stackrel{H_0}{=} 0$ and we can base
a test of $H_0$ on
\begin{eqnarray}
D_3|(\u S = \u s)\nonumber\\
\eqntext{\mbox{which has a known, computable distribution
under $H_0$.}}
\end{eqnarray}
This null distribution is known because $D_3$ has form $D_3 = D_3(\u
T)$ and assumption $B_2$ tells us that the distribution of $\u T|(\u S
= \u s)$ is
known. It is computable because under $B_1$, the distribution of
$D_3|(\u S = \u s)$ depends only on $\u y[1.\u s, 2.\u s]$ and under
$H_0^{RUs}$ the observed data $\u y[\u t.\u s]$ determines the
collection $\u y[1.\u s, 2.\u s]$.

An exact two-sided $p$-value can be computed as
\begin{eqnarray*}
&&\operatorname{pval}(D_{3,\mathrm{obs}})\\
&&\quad= P_{H_0}(|D_3|
\geq|D_{3,\mathrm{obs}}| | \u S = \u s)\\
&&\quad = P_{H_0}\bigl(\u T \in\bigl\{\u x:
|D_3(\u x)| \geq|D_{3,\mathrm{obs}}|\bigr\} | \u S = \u s\bigr).
\end{eqnarray*}
If we assume that $B_1$ and $B_2$ hold, an exact $p$-value $\leq\alpha$ gives
statistical evidence against $H_0^{RUs}$: $\u y[1.\u s] = \u y[2.\u s]$,
that is, evidence at the $\alpha$ level that $\u y[1.s_j] \neq \u
y[2.s_j]$ for at least one subject $s_j$. This test is called a Fisher
randomization test because it is based on the randomization approach
and it was described by \citet{FIS35}.

This Fisher randomization test based on $D_3$ is tailored to detect
differences between $\overline y[1.\u s]$ and $\overline y[2.\u s]$. To
detect other differences, such as scale differences between the $\u
y[1.\u s]$ and $\u y[2.\u s]$, an alternative to $D_3$ should (and can
easily) be used.

Attractive features of this Fisher randomization test include the
following: it has size guaranteed to be no larger than $\alpha$; it is
valid when the sampling depends on the process ($\u S \not\perp\!\!\!
\perp\u Y$); it does not require a model for the process variables $\u
Y$; and it does not require an estimate of the variance, $\operatorname{var}(D_3|\u S
= \u s)$.

\textit{Randomization vs. Permutation $P$-values}: It is clear
that this Fisher randomization test is conceptually very different from
the process-based permutation test. Indeed, as a rule, the
randomization $p$-value based on $D_3$ is numerically different than the
permutation $p$-value based on $D_1$. In fact, even if we had based both
$p$-values on the same statistic $D_1$, the $p$-values would generally be
different. There is an exception to this rule. Consider the special
case uniform randomization distribution,
%
\begin{equation}
P(\u T = \u x|\u S = \u s) = { \i(\u x \in \Pi(\u t))\over
{n \choose n_1}},\label{uniformT}
\end{equation}
where $n_1$ is the number of $1$'s in $\u t$ and $\Pi(\u t)$ is the set
of all rearrangements of $n_1$ $1$'s and $n_2=n-n_1$ $2$'s. This is the
randomization used in the special case two-treatment \emph{completely
randomized design} (e.g., Cox, \citeyear{Cox92}, pages~71--72; \citett{Kem77}, Section~8).
In this case, $D_1$ and $D_3$ are numerically identical, and the
randomization and permutation $p$-values are numerically identical. It is
this identity that often leads practitioners to incorrectly conclude
that the process-based permutation test is identical to the
randomization test. See \citet{Ern04} for an interesting discussion.

\subsubsection{Neyman randomization test}\label{sec6.3.2}

Compared to the view attributed to Fisher, Neyman was more interested
in detecting nonzero treatment effects of the aggregate variety,
especially $\overline y[1.\u s] - \overline y[2.\u s]$. He apparently
found it less practically useful to detect unit-specific effects if the
average effect was 0. For this reason, Neyman used the
no-average-effect hypothesis (cf. \citett{WEL37}). That is, Neyman viewed
the no-treatment-effect hypothesis as
\[
H_{0}^{RAs}: \overline y[1.\u s] = \overline y[2.\u s].
\]
Because $H_{0}^{RAs} \supset H_{0}^{RUs}$, Neyman's approach focused
on a narrower set of alternatives than Fisher, thereby opening up the
possibility of finding a test with higher power than the Fisher
randomization test, at least
for alternatives of practical (in Neyman's view) interest.

When $H_0 = (B_1, B_2, H_0^{RAs})$   holds, we have by (\ref{ED})
that $E(D_3|\u S = \u s) \stackrel{H_0}{=} 0$ and we can consider basing
a test of $H_0$ on
\begin{eqnarray*}
&&D_3|(\u S = \u s)\\
&&\quad\mbox{which has a known, but \emph{
noncomputable},}\\
&&\quad \mbox{distribution under $H_0$.}
\end{eqnarray*}
This null distribution is known because $D_3 = D_3(\u T)$ and $\u T|(\u
S = \u s)$ has a known distribution. It, however, is not
computable because it depends on $\u y[1.\u s, 2.\u s]$, which is not
determined by the observed data $\u y[\u t.\u s]$
under the no-average-effect hypothesis $H_0^{RAs}$. In contrast, recall
that under the more restrictive unit-specific (or sharp) null
$H_0^{RUs}$, the observed data did determine $\u y[1.\u s, 2.\u s]$.

Neyman was clearly aware of this noncomputability issue and instead
invoked a central limit theorem and used
\begin{eqnarray}
Z_3 \equiv Z_3(\u T) \equiv {D_3 \over \mathit{SE}(D_3|\u S = \u s)}
\nonumber\\
\eqntext{\mbox {where } \displaystyle Z_3|(\u S = \u s) \stackrel{H_0} {\sim}
\operatorname{approx} N(0,1).}
\end{eqnarray}
Here, $\mathit{SE}$ is a standard error, which is an estimator of the standard
deviation, $\mathrm{sd}(D_3  |  \u S = \u s)$. The standard deviation can be
computed using sampling theory as described in S\"arndal et al. (\citeyear{SarSweWre92}).
However, finding a reasonable estimator $\mathit{SE}$ of this standard deviation
is more difficult because of the 0 second-order inclusion
probabilities. Toward this end, Neyman (\citeyear{Ney90}) derived a reasonable
estimator of a tight upper bound for the variance under simplifying
assumptions on the inclusion probabilities (\citett{Rub90,Gad01};
see \citett{Cop73}, for a related result). It is useful to note that the
variance attains this upper bound when unit-treatment additivity holds,
that is, $\u y[1.s_j] = \u y[2.s_j] + \mathrm{constant},  j=1,\ldots, n$. In
this paper, we use the Neyman estimator of variance. The square root of
this estimator is $\mathit{SE}(D_3|\u S = \u s)$.

\begin{remark*}There is a related approximate normality result when
$H_0^{RAs}$ does not hold. Under $(B_1, B_2)$, we noted in (\ref{ED})
that $E(D_3|\u S = \u s) = \overline y[1.\u s] - \overline y[2.\u s]$
and we have that $\mathrm{sd}(D_3|\u S = \u s)$ is approximated by $\mathit{SE}(D_3|\u S
= \u s)$. By the central
limit theorem and continuous mapping results, we have
\[
{D_3 - (\overline y[1.\u s] - \overline y[2.\u s]) \over \mathit{SE}(D_3|\u S =
\u s)} | (\u S = \u s) \sim \operatorname{approx} N(0,1).
\]
This result is useful for testing other hypotheses and for computing
confidence intervals.
\end{remark*}

The Normal approximation for $Z_3$ generally improves as the number of
support points in $\u T|(\u S = \u s)$ increases. However, when the
differences $\u y[1.s_j] - \u y[2.s_j]$ are highly variable, the unit
variance in the approximation can be a substantial overestimate (see
\citett{Gad01}), and when $\u y[1.s_j] - \u y[2.s_j] = \mathrm{constant}$, the
unit variance can be a slight underestimate when the sample sizes are
small (based on observations from the simulation study carried out for
this paper).

An approximate two-sided $p$-value can be computed as
\begin{eqnarray*}
P_{H_0}\bigl(|Z_3| \geq|Z_{3,\mathrm{obs}}| | \u S = \u s\bigr)
&\approx& P\bigl(\bigl|N(0,1)\bigr| \geq|Z_{3,\mathrm{obs}}|\bigr)\\
& \equiv& \mathrm{apval}(Z_{3,\mathrm{obs}}).
\end{eqnarray*}
If $B_1$ and $B_2$ hold, an approximate $p$-value $\leq\alpha$ gives
statistical evidence against $H_0^{RAs}$: $\overline y[1.\u s] =
\overline y[2.\u s]$, that is, evidence at the approximate $\alpha$
level that $\overline y[1.\u s] \neq \overline y[2.\u s]$. This test
is called a Neyman randomization test because it is based on the
randomization approach and ideas in Neyman (\citeyear{Ney90}).

Unlike the Fisher randomization test of $H_{0}^{RUs}$, the size of the
Neyman test of $H_{0}^{RAs}$ is not guaranteed to be less than or equal
to $\alpha$; it is only approximately size $\alpha$. For smaller $n_1$
and $n_2$ and when the more restrictive hypothesis $H_0^{RUs}$ holds,
the Neyman randomization test tends to be anti-conservative, with size
a bit larger than the nominal $\alpha$. This follows because the Neyman
estimator of the variance tends to slightly underestimate the true
variance in this case. For moderate $n_1$ and $n_2$ the approximation
is usually reasonable provided $D_3(\u T)$ has enough support points
with respect to the $\u T|(\u S = \u s)$ distribution. We empirically
explore this approximation below.

\subsection{Selection-Based Tests}\label{sec6.4}

With the selection-based approach, we condition on the process values
[only $(\u S, \u T)$ is random] and use
\[
\u y[\u t.\u s] \leftarrow \u Y[\u T.\u S] | (\u Y = \u y) \sim \u y[\u T.\u S]
| (\u Y = \u y)
\]
to carry out inferences about $\u y[1.\u P, 2.\u P]$. The
selection-based test procedures outlined below are valid provided the
assumptions $C_1$ and $C_2$ of Section~\ref{sec5.3} hold. There it was pointed
out that these two assumptions are often untenable, so the following
test procedures must be applied with caution.

\subsubsection{Fisher selection test}\label{sec6.4.1}

The no-unit-specific-treatment-effect (or sharp null) hypothesis in
this selection-based setting has the form
$
H_{0}^{RUP}$: $\u y[1.i] = \u y[2.i],   i=1,\ldots, N$, or, more simply,
\[
H_0^{RUP}: \u y[1.\u P] = \u y[2.\u P].
\]
When $H_0 = (C_1, C_2, H_0^{RUP})$   holds, we have by (\ref{ED})
that $E(D_{23}) \stackrel{H_0}{=} 0$ and we can consider basing
a test of $H_0$ on
\begin{eqnarray*}
&&D_{23}|(\u S = \u s)\\
&&\quad \mbox{which has a known, but \emph
{noncomputable}}\\
&&\quad\mbox{distribution under $H_0$.}
\end{eqnarray*}
This null distribution is known because $D_{23}$ has form $D_{23} =
D_{23}(\u S,\u T)$ and assumption $C_2$ tells us that the distribution
of $(\u S,\u T)$ is known. It is, however, not computable because it
depends on $\u y[1.\u P, 2.\u P]$, which is not determined by the
observed data $\u y[\u t.\u s]$ under the hypothesis $H_0^{RUP}$. To
see this, note that for $\u s' \neq \u s$, there
is an $s'_j$ such that both $\u y[1.s'_j]$ and $\u y[2.s'_j]$ are
unobserved and hence not computable even under $H_0^{RUP}$.

It follows that an exact Fisher selection test is \textit{not}
available in this selection-based setting. We could condition on the
sample and be content using the Fisher randomization test to draw
inferences about $\u y[1.\u s, 2.\u s]$ rather than $\u y[1.\u P, 2.\u
P]$. Alternatively, we could use the approximate selection-based test
described in the next subsection.

\subsubsection{Neyman selection test}\label{sec6.4.2}

In analogy to the randomization setting, Neyman likely would consider
the no-average-effect hypothesis:
\[
H_0^{RAP}: \overline y[1.\u P] = \overline y[2.\u P].
\]
When $H_0 = (C_1, C_2, H_0^{RAP})$   holds, we have by (\ref{ED})
that $E(D_{23}) \stackrel{H_0}{=} 0$ and, analogous to the
randomization setting, we can base a test of $H_0$ on
\begin{eqnarray}
Z_{23} \equiv Z_{23}(\u S, \u T) =
{D_{23}\over
\mathit{SE}(D_{23})}\nonumber\\
\eqntext{\mbox{where } \displaystyle Z_{23} \stackrel{H_0}
{\sim} \operatorname{approx} N(0,1).}
\end{eqnarray}
Just as with $\mathrm{sd}(D_3|\u S= \u s)$ in the randomization approach, the
standard deviation $\mathrm{sd}(D_{23})$ can be computed and estimated using
sampling theory. The estimation, however, is subject to the same
problems as in the randomization approach because of the
nonmeasurability of probability sample $\u T.\u S$. Suffice it to say
that a reasonable Neyman estimator $\mathit{SE}(D_{23})$, analogous to the one
in the randomization setting, exists.

The approximate Normality result follows just as in the randomization
setting. Specifically, under $C_1$ and $C_2$, and using the same
arguments as in the randomization approach, we have that quite generally
\[
{D_{23} - (\overline y[1.\u P] - \overline y[2.\u P])\over \mathit{SE}(D_{23})} \sim \operatorname{approx} N(0,1).
\]
The approximation generally improves as the number of support points in
$\u T.\u S$ increases. However, when the differences $\u y[1.i] - \u
y[2.i]$ are highly variable, the unit variance in the approximation can
be a substantial overestimate (see \citett{Gad01}).

An approximate two-sided $p$-value can be computed as
\begin{eqnarray*}
P_{H_0}\bigl(|Z_{23}| \geq|Z_{23,\mathrm{obs}}| \bigr) &\approx& P
\bigl(\bigl|N(0,1)\bigr| \geq |Z_{23,\mathrm{obs}}|\bigr) \\
&\equiv& \mathrm{apval}(Z_{23,\mathrm{obs}}).
\end{eqnarray*}
If $C_1$ and $C_2$ hold, an approximate $p$-value $\leq\alpha$ gives
statistical evidence against $H_0^{RAP}$: $\overline y[1.\u P] =
\overline y[2.\u P]$, that is, evidence at the approximate $\alpha$
level that $\overline y[1.\u P] \neq \overline y[2.\u P]$. This test
is called a Neyman selection test because it is based on the selection
approach and ideas in Neyman (\citeyear{Ney90}).

Just as in the randomization setting, the size of the Neyman test of
$H_{0}^{RAP}$ is not guaranteed to be less than or equal to $\alpha$;
it is only approximately size $\alpha$. Remarks regarding the
approximation in this selection setting are analogous to those given at
the end of Section~\ref{sec6.3.2}, in the randomization setting.

\section{Empirical Investigations}\label{sec7}

\subsection{Cell Phone Use Example (Revisited)}\label{sec7.1}
The process variable $\u Y[t.i]$ is defined as the reaction time for
the $i$th unit in population $\u P$ when exposed to treatment $t$.
Inference about the process $\u Y$ distribution will be difficult to
describe because the sample of $64$ students was not taken from any
well-defined population $\u P$. For any substantively interesting
population, for example, $\u P = $ licensed drivers in Utah, the assumption
that $\u S \perp\!\!\!\perp \u Y$ is untenable given the haphazard
nature of the sample selection. The untenability of $\u S \perp\!\!\!
\perp\u Y$ also implies that it will be difficult to carry out
inferences about the population values $\u y[1.\u P, 2.\u P]$ for any
substantively interesting population $\u P$. For these reasons, it
makes sense to focus on inferences about the 128 potential values in
$\u y[1.\u s, 2.\u s]$. That is, it is arguably better to use
randomization-based inference for this example.

We assume that the randomization was carried out mechanically so that
$\u T \perp\!\!\!\perp\u Y|\u S$ and we assume that the distribution
of $\u T|(\u S = \u s)$ is uniform in
the sense of (\ref{uniformT}); that is, conditions $B_1$ and $B_2$ of
Section~\ref{sec6.3} are assumed to hold. We will use the Fisher randomization
test to test the no-treatment-effect hypothesis $ H_0^{RUs}$: $\u
y[1.s_j] = \u y[2.s_j],  j=1,\ldots, 64$ and the Neyman randomization
test to test the no-treatment-effect hypothesis $H_0^{RAs}$: $\overline
y[1.\u s] = \overline y[2.\u s]$.

For these data, the observed randomization statistics are
\begin{eqnarray*}
D_{3,\mathrm{obs}} &= &51.59,\quad Z_{3,\mathrm{obs}} = {51.59\over19.30} =
2.67,\\
\operatorname{pval}(D_{3,\mathrm{obs}}) &=& 0.0074\quad \mbox{and} \\
\mathrm{apval}(Z_{3,\mathrm{obs}})& = &0.0075.
\end{eqnarray*}
Because the Fisher randomization $p$-value\break $\operatorname{pval}(D_{3,\mathrm{obs}}) = 0.0074$ is
small, we have sufficient evidence to
reject $H_0^{RUs}$; there is statistical evidence that $\u y[1.s_j] \neq \u y[2.s_{j}]$ for at least one subject in the sample of 64.
Because the Neyman\vadjust{\goodbreak} randomization $p$-value $\mathrm{apval}(Z_{3,\mathrm{obs}}) = 0.0075$ is
small, we have sufficient evidence to
reject $H_0^{RAs}$; there is statistical evidence that $\overline
y[1.\u s] \neq \overline y[2.\u s]$. In fact, because
$D_{3,\mathrm{obs}} = 51.59$ is a Horvitz--Thompson unbiased estimate of
$\overline y[1.\u s] - \overline y[2.\u s]$, the Neyman test gives
statistical evidence that the reaction time values are higher on
average when cell phones are used, at least for this sample of 64. In
other words, there is statistical evidence of a treatment effect.

For completeness and for comparison purposes, we also give the values
of the other commonly used $p$-values, viz., permutation, Wilcoxon,
Welch's approximate $t$, and the pooled $t$:
\begin{eqnarray*}
\operatorname{pval}(D_{1,\mathrm{obs}}) &=& 0.0074,\quad \operatorname{pval}
(W_{\mathrm{obs}}) = 0.0184 ,\\
\mathrm{apval}(T_{\mathrm{obs}}) &=& 0.0110 \quad\mbox{and}\quad \operatorname{pval}(T_{p,\mathrm{obs}}) = 0.0107.
\end{eqnarray*}
Strictly speaking, these are only applicable for process-based
inference, so they are of questionable utility for this example.
As noted above, because the randomization distribution is uniform, the
permutation $p$-value $\operatorname{pval}(D_{1,\mathrm{obs}})$ is numerically (but not
conceptually!) identical to the Fisher randomization $p$-value $\operatorname{pval}(D_{3,\mathrm{obs}})$.

All the computations were carried out in R. The author has written code
to compute the Neyman randomization $p$-value. The Fisher randomization
and permutation $p$-values were approximated using Monte-Carlo estimation
(here we used $10^6$ simulations) as carried out in
\texttt{twot.permutation $\{$DAAG$\}$}. The Wilcoxon $p$-value was computed using
\texttt{wilcox.test $\{$stats$\}$}. Note that when there are ties, as
there are in this example, \texttt{wilcox.test} only reports approximate $p$-values.

\subsection{A Simulation Study}\label{sec7.2}

This section empirically compares the operating characteristics of the
different tests considered in this paper, under a variety of scenarios.
All computations were carried out in R, with $p$-values computed as
described at the end of the previous subsection. The simulated data are
generated according to models of the form
%
\begin{eqnarray}\label{dgm_simulation}
\u y[1.i] &\leftarrow& \u Y[1.i]\ \mathrm{IID} \sim
[\mathrm{scenario}],\nonumber
\\
\u y[2.i] &\leftarrow&\u Y[2.i] \sim [\mathrm{scenario}],\qquad i=1,\ldots, N,
\nonumber\\
\u s &\leftarrow&\u S|(\u Y = \u y) \sim P\bigl(\u S = (1,\ldots,n) | \u Y = \u
y\bigr) = 1,
\nonumber
\\[-8pt]
\\[-8pt]
 \eqntext{\mbox{where $n = N$},}
\\
\u t &\leftarrow&\u T|(\u Y = \u y, \u S = \u s) \sim P\bigl(\u T = \u
t'| \u Y = \u y, \u S = \u s\bigr)\nonumber\\
& =&
{n_1! n_2!\over n!}\i \bigl(\u t' \in {\cal T}\bigr),\nonumber
\end{eqnarray}
where $n = n_1 + n_2$ and ${\cal T}$ is the set of all possible
rearrangements of $n_1$ $1$'s and $n_2$ $2$'s. Looking back at the
process-based assumptions of
Section~\ref{sec5.1}, we see that $A_1$ holds, but none of $A_2$--$A_7$ is
guaranteed to hold. Both the randomization-based assumptions $B_1$ and
$B_2$ of Section~\ref{sec5.2} hold, as do both the selection-based assumptions
$C_1$ and $C_2$ of Section~\ref{sec5.3}. A more extensive simulation would
also investigate scenarios where more of the assumptions do not hold.

For data-generation models of the form (\ref{dgm_simulation}), we have
that (i) the randomization- and selection-based approaches are
identical because the sample $\u S$ is taken to be equal to the
population $\u P$ with probability one; and (ii) the permutation and
Fisher randomization $p$-values are numerically (not conceptually!)
identical because the randomization distribution (the distribution of
$\u T$) is uniform over the set of all possible treatment assignments.

Although the permutation-, Wilcoxon-, and $t$-tests are process-based
approaches, we will estimate their operating characteristics for both
the process and randomization (here, randomization${}={}$selection) distributions.
Similarly, the Fisher and Neyman randomization tests are
randomization-based approaches, but we report their operating
characteristics for both the process and the randomization distributions.
In the tables below, the rows labeled ``Randomization'' give Monte-Carlo
estimates of the power of the tests over the distribution $\u T|(\u Y =
\u y, \u S = \u s)$.
The rows labeled ``Process'' give Monte-Carlo estimates of the power of
the tests over the distribution $\u Y|(\u S = \u s, \u T = \u t)$. In
all cases, the nominal size is set at $\alpha= 0.05$.

\begin{table*}[t]
\caption{Monte-Carlo estimates of size when $n_1 =
n_2 = 10$, nominal size${}={}$5\%}\label{tab3}
\begin{tabular*}{\textwidth}{@{\extracolsep{\fill}}lccccccc@{}}
\hline
$\bolds{n_1=n_2 = 10}$ &\textbf{Permutation}\tabnoteref{tt1} &
\textbf{Wilcoxon} & \textbf{$\bolds{t}$(Welch)} & \textbf{$\bolds{t}$(Pooled)} &\textbf{Fisher\tabnoteref{tt1}} & \textbf{Neyman} \\
\hline
$H_0^{UP}$ &\multicolumn{6}{l}{$ \u y[1.i] \leftarrow\u Y[1.i] \ \mathrm{IID}
 \sim N(10, 2^2)$}& Sc.~1\\
true&\multicolumn{6}{l}{$\u y[2.i] \leftarrow\u Y[2.i] = \u Y[1.i],   i=1,\ldots, 20$}&\\[3pt]
Randomization &4.6& 3.6 &4.7 & 4.7 &4.6 &6.5 \\
Process &4.3& 3.4 &4.2 & 4.3 & 4.3 & 6.9 \\[6pt]
$H_0^{UP}$ &\multicolumn{6}{l}{$ \u y[1.i] \leftarrow\u Y[1.i] \ \mathrm{IID}
 \sim \operatorname{Gamma}(\mathrm{shape}=1,\mathrm{scale}=5)$}&Sc.~2\\
true&\multicolumn{6}{l}{$\u y[2.i] \leftarrow\u Y[2.i] = \u Y[1.i],   i=1,\ldots, 20$} & \\[3pt]
Randomization &5.0& 4.9 &4.1 & 4.6 &5.0 &7.4\\
Process &4.0& 4.1 &3.2 & 3.5 & 4.0 &7.7 \\[6pt]
$H_0^{UP}$ &\multicolumn{6}{l}{$ \u y[1.i] \leftarrow\u Y[1.i]\ \mathrm{IID}
 \sim 0.9 U(0,20) + 0.1 U(200,201)$, ``mixture of uniforms''}&Sc.~3\\
true&\multicolumn{6}{l}{$\u y[2.i] \leftarrow\u Y[2.i] = \u Y[1.i],   i=1,\ldots, 20$} & \\[3pt]
Randomization\tabnoteref{tt2} &4.6& 3.9 &0.0 & 0.0 &4.6 &0.0\\
Process &3.8 & 3.5 &1.1 & 1.8 & 3.8 & 11.2\\[6pt]
$ H_0^{EUP},  H_0^{RAs}$ &\multicolumn{6}{l}{$ \u y[1.i] \leftarrow
\u Y[1.i]\ \mathrm{IID}  \sim N(10, 2^2)$}&Sc.~4\\
true&\multicolumn{6}{l}{$\u y[2.i] \leftarrow\u Y[2.i] = \u Y[1.i] +
E_i - \overline E,   E_i \ \mathrm{IID} \sim N(0, 3^2),   i=1,\ldots,
20$} & \\[3pt]
Randomization &1.5&1.9 &1.7 &1.8 &1.5 &3.3 \\
Process &2.7&2.0 &2.5&2.6&2.7&4.2 \\[6pt]
$H_0^{EUP},  H_0^{RAs}$ &\multicolumn{6}{l}{$ \u y[1.i] \leftarrow\u
Y[1.i] \ \mathrm{IID}  \sim \operatorname{Gamma}(\mathrm{shape}=1,\mathrm{scale}=5)$}&Sc.~5\\
true&\multicolumn{6}{l}{$\u y[2.i] \leftarrow\u Y[2.i] = 2 \u Y[1.i]
- \overline Y[1.\u P],   i=1,\ldots, 20$} & \\[3pt]
Randomization &4.8& 6.8 &4.3 &4.4 &4.8 &7.6 \\
Process &4.0& 7.4 & 3.6 &3.7 &4.0 &7.4 \\[6pt]
$H_0^{UP}$ &\multicolumn{6}{l}{$ \u y[1.i] \leftarrow\u Y[1.i] \ \mathrm{IID}
 {\sim}  \operatorname{bin}(1,0.28)$}&Sc.~6\\
true&\multicolumn{6}{l}{$\u y[2.i] \leftarrow\u Y[2.i] = \u Y[1.i],   i=1,\ldots, 20$}& \\[3pt]
Randomization\tabnoteref{tt3} &0.0&NA&9.1&9.1&0.0&9.1 \\
Process & 2.1&NA &4.5&4.5&2.1&11.3 \\[6pt]
$H_0^{DUP},  H_0^{RAs}$ &\multicolumn{6}{l}{$ \u y[1.i] \leftarrow\u
Y[1.i] \ \mathrm{IID}   {\sim}\tabnoteref{tt4}  \operatorname{bin}(1,0.28)$}&Sc.~7\\
true&\multicolumn{6}{l}{$\u y[2.i] \leftarrow\u Y[2.i] \ \mathrm{IID}  {\sim
}\tabnoteref{tt4}  \operatorname{bin}(1,0.28),  \operatorname{corr}(\u Y[1.i],\u Y[2.i]) = 0.37,   i=1,\ldots,
20$}& \\[3pt]
Randomization\tabnoteref{tt5} &0.4&NA\tabnoteref{tt6} &1.6 &1.6&0.4&4.6\\
Process &0.2& NA&1.0&1.0&0.2&3.7 \\
\hline
\end{tabular*}
\tabnotetext[]{}{Table entries give the rejection rates (as a percent) for the 1000 simulations.}
\tabnotetext[]{}{All indented hypotheses are also true; see Section~\ref{sec4.2}. For example, in
row 1, $H_0^{UP}$ is true. It follows that all the other hypotheses in
Section~\ref{sec4.2} are also true.}
\tabnotetext[a]{tt1}{For this simulation, the permutation and Fisher randomization test
results are numerically identical.}
\tabnotetext[b]{tt2}{The fixed $\u y$ includes one large observation from the
$U(200,201)$ distribution.}
\tabnotetext[c]{tt3}{The fixed $\u y[1.\u P] = 0\ 0\ 0\ 0\ 0\ 1\ 0\ 0\ 0\ 0\ 0\ 0\ 0\ 0\ 0\ 1\ 1\ 0\ 1\
0 = \u y[2.\u P]$.}
\tabnotetext[d]{tt4}{This is an approximation because the $\u Y$ values are adjusted to
satisfy $H_0^{RAs}$.}
\tabnotetext[e]{tt5}{The fixed $\u y[1.\u P] = 1\ 0\ 0\ 1\ 0\ 1\ 0\ 1\ 0\ 0\ 1\ 0\ 0\ 1\ 0\ 0\ 0\ 0\ 0\
0$, $\u y[2.\u P] = 0\ 0\ 0\ 0\ 1\ 1\ 0\ 1\ 0\ 0\ 1\ 0\ 0\ 1\ 1\ 0\ 0\ 0\ 0\ 0$.}
\tabnotetext[f]{tt6}{Because of the many ties in the binomial case, the Wilcoxon test
as described herein is not applicable.}
\end{table*}

\begin{table*}
\caption{Monte-Carlo estimates of power when $n_1 =
n_2 = 10$, nominal size${}={}$5\%}\label{tab4}
\begin{tabular*}{\textwidth}{@{\extracolsep{\fill}}lccccccc@{}}
\hline
$\bolds{n_1=n_2 = 10}$ &\textbf{Permutation}\tabnoteref{ttt1} &
\textbf{Wilcoxon} & \textbf{$\bolds{t}$(Welch)} & \textbf{$\bolds{t}$(Pooled)} &\textbf{Fisher\tabnoteref{ttt1}} & \textbf{Neyman} \\
\hline
$ H_0^{EUP},  H_0^{RAs} $ &\multicolumn{6}{l}{$ \u y[1.i] \leftarrow
\u Y[1.i] \ \mathrm{IID}  \sim N(10, 2^2)$}&Sc.~1\\
false&\multicolumn{6}{l}{$\u y[2.i] \leftarrow\u Y[2.i] = \u Y[1.i] +
2 ,   i=1,\ldots, 20$}&\\[3pt]
Randomization &52.7&49.3 &51.3& 52.5 &52.7 &59.9 \\
Process &55.9& 51.6 & 55.5 &56.1 & 55.9 &62.7 \\[6pt]
$H_0^{EUP},  H_0^{RAs}$ &\multicolumn{6}{l}{$ \u y[1.i] \leftarrow\u
Y[1.i] \ \mathrm{IID}  \sim N(10, 2^2)$}&Sc.~2\\
false &\multicolumn{6}{l}{$\u y[2.i] \leftarrow\u Y[2.i] = \u Y[1.i]
+ 2 + E_i - \overline E,   E_i \ \mathrm{IID} \sim N(0, 3^2),   i=1,\ldots
, 20$}&\\[3pt]
Randomization &26.2& 23.6 & 24.1 &25.7 &26.2 &35.6 \\
Process &28.6& 23.8 &26.1 &27.2 &28.6 &36.6 \\[6pt]
$H_0^{EUP},  H_0^{RAs}$ &\multicolumn{6}{l}{$ \u y[1.i] \leftarrow\u
Y[1.i]\ \mathrm{IID}  \sim N(10, 2^2)$}&Sc.~3\\
false&\multicolumn{6}{l}{$\u y[2.i] \leftarrow\u Y[2.i] = 1.2 \u
Y[1.i],   i=1,\ldots, 20$}&\\[3pt]
Randomization & 34.7 &27.1 &34.8 &35.3 &34.7 &43.0 \\
Process &48.4 &43.0 &47.5 &48.4 &48.4 &57.2 \\[6pt]
$H_0^{EUP},  H_0^{RAs}$ &\multicolumn{6}{l}{$ \u y[1.i] \leftarrow\u
Y[1.i] \ \mathrm{IID}  \sim \operatorname{Gamma}(\mathrm{shape}=1,\mathrm{scale}=5)$}&Sc.~4\\
false&\multicolumn{6}{l}{$\u y[2.i] \leftarrow\u Y[2.i] = 2 \u
Y[1.i],  i=1,\ldots, 20$}&\\[3pt]
Randomization & 19.2 &12.9 &16.1 &18.7 &19.2 &28.6 \\
Process &30.2 & 23.7 &23.4 &26.0 &30.2 &38.5 \\[6pt]
$H_0^{EUP},  H_0^{RAs}$ &\multicolumn{6}{l}{$ \u y[1.i] \leftarrow\u
Y[1.i] \ \mathrm{IID}  \sim \operatorname{Gamma}(\mathrm{shape}=1,\mathrm{scale}=5)$}&Sc.~5\\
false &\multicolumn{6}{l}{$\u y[2.i] \leftarrow\u Y[2.i] = 3 \u
Y[1.i] + E_i, E_i \ \mathrm{IID}  \sim N(0,5^2),   i=1,\ldots, 20$}&\\[3pt]
Randomization &45.7 &28.6 &40.2 &45.3 &45.7 &65.5 \\
Process &49.2& 38.1 &39.8 &44.4 &49.2 &63.5 \\[6pt]
$H_0^{EUP},  H_0^{RAs}$ &\multicolumn{6}{l}{$ \u y[1.i] \leftarrow\u
Y[1.i] \ \mathrm{IID}  {\sim}  \operatorname{bin}(1,0.28)$}&Sc.~6\\
false&\multicolumn{6}{l}{$\u y[2.i] \leftarrow\u Y[2.i] \ \mathrm{IID}   {\sim
}  \operatorname{bin}(1,0.71),  \operatorname{corr}(\u Y[1.i],\u Y[2.i]) = 0.29,   i=1,\ldots,
20$}&\\[3pt]
Randomization\tabnoteref{ttt2}&18.9 &NA&37.3&37.3&18.9&37.4 \\
Process &29.6& NA&48.0&48.0&29.6&50.3 \\
\hline
\end{tabular*}
\tabnotetext[]{}{Table entries give the rejection rates (as a percent) for the 1000 simulations.}
\tabnotetext[a]{ttt1}{For this simulation, the permutation and Fisher randomization test
results are numerically identical.}
\tabnotetext[b]{ttt2}{The fixed $\u y[1.\u P] = $ 0 0 0 0 0 1 0 0 1 0 1 0 0 0 0 1 1 0 1
0,  $\u y[2.\u P] = $ 0 1 1 1 1 1 1 0 1 0 1 1 0 0 1 1 1 0 1 1.}
\end{table*}

\begin{table*}
\caption{Monte-Carlo estimates of size when $n_1 =
n_2 = 50$, nominal size${}={}$5\%}\label{tab5}
\begin{tabular*}{\textwidth}{@{\extracolsep{\fill}}lccccccc@{}}
\hline
$\bolds{n_1=n_2 = 50}$ &\textbf{Permutation}\tabnoteref{tttt1} &
\textbf{Wilcoxon} & \textbf{$\bolds{t}$(Welch)} & \textbf{$\bolds{t}$(Pooled)} &\textbf{Fisher\tabnoteref{tttt1}} & \textbf{Neyman} \\
\hline
$H_0^{UP}$ &\multicolumn{6}{l}{$ \u y[1.i] \leftarrow\u Y[1.i]\ \mathrm{IID}
 \sim N(10, 2^2)$}&Sc.~1\\
true&\multicolumn{6}{l}{$\u y[2.i] \leftarrow\u Y[2.i] = \u Y[1.i],   i=1,\ldots, 100$}& \\[3pt]
Randomization &4.0&4.0 &4.0 &4.0 &4.0 &4.4 \\
Process &4.7&4.8 &4.8 &4.8 &4.7&5.5 \\[6pt]
$H_0^{UP}$ &\multicolumn{6}{l}{$ \u y[1.i] \leftarrow\u Y[1.i]\ \mathrm{IID}
 \sim \operatorname{Gamma}(\mathrm{shape}=1,\mathrm{scale}=5)$}&Sc.~2\\
true&\multicolumn{6}{l}{$\u y[2.i] \leftarrow\u Y[2.i] = \u Y[1.i],   i=1,\ldots, 100$}& \\[3pt]
Randomization &4.9& 5.0 & 4.8&4.8 &4.9 &5.4\\
Process &4.1& 3.9 &3.9&3.9 &4.1 &4.8\\[6pt]
$H_0^{UP}$ &\multicolumn{6}{l}{$ \u y[1.i] \leftarrow\u Y[1.i]\ \mathrm{IID}
 \sim 0.9 U(0,20) + 0.1 U(200,201)$, ``mixture of uniforms''}&Sc.~3\\
true &\multicolumn{6}{l}{$\u y[2.i] \leftarrow\u Y[2.i] = \u Y[1.i],
  i=1,\ldots, 100$}& \\[3pt]
Randomization\tabnoteref{tttt2}&4.2&6.5&4.3&4.5&4.2&8.6 \\
Process &5.3&5.4&5.3&5.3&5.3&6.6 \\[6pt]
$ H_0^{EUP},  H_0^{RAs}$ &\multicolumn{6}{l}{$ \u y[1.i] \leftarrow
\u Y[1.i] \ \mathrm{IID}  \sim N(10, 2^2)$}&Sc.~4\\
true&\multicolumn{6}{l}{$\u y[2.i] \leftarrow\u Y[2.i] = \u Y[1.i] +
E_i - \overline E,   E_i \ \mathrm{IID} \sim N(0, 3^2),   i=1,\ldots,
100$}& \\[3pt]
Randomization &2.5& 3.1 &2.4 &2.4 &2.5 &3.4 \\
Process &3.0& 4.6 & 2.9 &3.2 &3.0 &3.9 \\[6pt]
$H_0^{EUP},  H_0^{RAs}$ &\multicolumn{6}{l}{$ \u y[1.i] \leftarrow\u
Y[1.i] \ \mathrm{IID}  \sim \operatorname{Gamma}(\mathrm{shape}=1,\mathrm{scale}=5)$} &Sc.~5\\
true&\multicolumn{6}{l}{$\u y[2.i] \leftarrow\u Y[2.i] =2 \u Y[1.i] -
\overline Y[1,\u P],     i=1,\ldots, 100$}&\\[3pt]
Randomization &4.6&42.5 &4.4 &4.4&4.6&6.1 \\
Process &2.8&35.1 &2.8 &2.8 &2.8 &5.0 \\[6pt]
$H_0^{UP}$&\multicolumn{6}{l}{$ \u y[1.i] \leftarrow\u Y[1.i] \ \mathrm{IID}
  {\sim}  \operatorname{bin}(1,0.28)$}&Sc.~6\\
true&\multicolumn{6}{l}{$\u y[2.i] \leftarrow\u Y[2.i] = \u Y[1.i],   i=1,\ldots, 100$}& \\[3pt]
Randomization\tabnoteref{tttt3} &2.2&NA&5.5&5.5&2.2&5.5 \\
Process &3.5& NA &5.0&5.0&3.5&5.9 \\[6pt]
$H_0^{DUP}, H_0^{RAs}$ &\multicolumn{6}{l}{$ \u y[1.i] \leftarrow
\u Y[1.i] \ \mathrm{IID}   {\sim}\tabnoteref{tttt4}  \operatorname{bin}(1,0.28)$}&Sc.~7\\
true&\multicolumn{6}{l}{$\u y[2.i] \leftarrow\u Y[2.i] \ \mathrm{IID}  {\sim
}\tabnoteref{tttt4}  \operatorname{bin}(1,0.28),  \operatorname{corr}(\u Y[1.i],\u Y[2.i]) = 0.37,   i=1,\ldots,
100$}& \\[3pt]
Randomization\tabnoteref{tttt5} &0.5 &NA&0.8&0.8&0.5&1.0 \\
Process &1.6& NA &2.8&2.8&1.6&3.5 \\
\hline
\end{tabular*}
\tabnotetext[]{}{Table entries give the rejection rates (as a percent) for the 1000 simulations.}
\tabnotetext[a]{tttt1}{For this simulation, the permutation and Fisher randomization test
results are numerically identical.}
\tabnotetext[b]{tttt2}{The fixed $\u y$ includes 7 large observations from the
$U(200,201)$ distribution.}
\tabnotetext[c]{tttt3}{The fixed $\u y[1.\u P] = \u y[2.\u P]$ with $\overline y[1.\u P]
= \overline y[2.\u P] = 32/100$.}
\tabnotetext[d]{tttt4}{This is an approximation because the $\u Y$ values are adjusted to
satisfy $H_0^{RAs}$.}
\tabnotetext[e]{tttt5}{The fixed $\u y$ is such that $\u y[1.\u P] \neq \u y[2.\u P]$,
  $\overline y[1.\u P] = \overline y[2.\u P] = 33/100$, and $\operatorname{corr}(\u
y[1.\u P], \u y[2.\u P]) = 0.186$.}
\end{table*}

\begin{table*}
\caption{Monte-Carlo estimates of power when $n_1 =
n_2 = 50$, nominal size${}={}$5\%}\label{tab6}
\begin{tabular*}{\textwidth}{@{\extracolsep{\fill}}lccccccc@{}}
\hline
$\bolds{n_1=n_2 = 50}$ &\textbf{Permutation}\tabnoteref{ttttt1} &
\textbf{Wilcoxon} & \textbf{$\bolds{t}$(Welch)} & \textbf{$\bolds{t}$(Pooled)} &\textbf{Fisher\tabnoteref{ttttt1}} & \textbf{Neyman} \\
\hline
$ H_0^{EUP},  H_0^{RAs} $ &\multicolumn{6}{l}{$ \u y[1.i] \leftarrow
\u Y[1.i] \ \mathrm{IID}  \sim N(10, 2^2)$}&Sc.~1\\
false&\multicolumn{6}{l}{$\u y[2.i] \leftarrow\u Y[2.i] = \u Y[1.i]
+ 1 ,   i=1,\ldots, 100$}& \\[3pt]
Randomization & 80.9& 76.4 &80.4 &80.4 &80.9 &81.3 \\
Process &69.5& 67.7 &69.9 &69.9 &69.5 &70.4 \\[6pt]
$H_0^{EUP},  H_0^{RAs}$&\multicolumn{6}{l}{$ \u y[1.i] \leftarrow\u
Y[1.i] \ \mathrm{IID}  \sim N(10, 2^2)$}&Sc.~2\\
false &\multicolumn{6}{l}{$\u y[2.i] \leftarrow\u Y[2.i] = \u Y[1.i]
+ 1 + E_i - \overline E,   E_i \ \mathrm{IID} \sim N(0, 3^2),  i=1,\ldots,
100$}& \\[3pt]
Randomization &36.3&31.4 &36.2 &36.4 &36.3 &42.7 \\
Process &37.9&36.3 &37.5&38.0&37.9&42.8 \\[6pt]
$H_0^{EUP},  H_0^{RAs}$&\multicolumn{6}{l}{$ \u y[1.i] \leftarrow\u
Y[1.i] \ \mathrm{IID} \sim N(10, 2^2)$}&Sc.~3\\
false &\multicolumn{6}{l}{$\u y[2.i] \leftarrow\u Y[2.i] = 1.1 \u
Y[1.i],   i=1,\ldots, 100$}&\\[3pt]
Randomization & 70.5& 68.6& 71.0 &71.1 &70.5 &72.1 \\
Process &66.6& 63.9 & 65.7 &65.7 &66.6 &67.4 \\[6pt]
$H_0^{EUP},  H_0^{RAs}$ &\multicolumn{6}{l}{$ \u y[1.i] \leftarrow\u
Y[1.i]\ \mathrm{IID}  \sim \operatorname{Gamma}(\mathrm{shape}=1,\mathrm{scale}=5)$}&Sc.~4\\
false&\multicolumn{6}{l}{$\u y[2.i] \leftarrow\u Y[2.i] = 1.5 \u
Y[1.i],   i=1,\ldots, 100$}&\\[3pt]
Randomization &46.6&39.0 &46.2 &46.4 &46.6 &49.5 \\
Process &49.2& 40.6 &48.0 &48.2 &49.2 &51.8 \\[6pt]
$H_0^{EUP},  H_0^{RAs}$ &\multicolumn{6}{l}{$ \u y[1.i] \leftarrow\u
Y[1.i]\  \mathrm{IID}  \sim \operatorname{Gamma}(\mathrm{shape}=1,\mathrm{scale}=5)$}&Sc.~5\\
false&\multicolumn{6}{l}{$\u y[2.i] \leftarrow\u Y[2.i] = 1.5 \u
Y[1.i] + E_i, E_i\ \mathrm{IID}  \sim N(0,5^2),   i=1,\ldots, 100$}&\\[3pt]
Randomization & 41.0 &35.2 &40.4 &40.5 &41.0 &44.3 \\
Process &39.2 &30.7 &38.9 &39.0 &39.2 &44.2 \\[6pt]
$H_0^{EUP},  H_0^{RAs}$ &\multicolumn{6}{l}{$ \u y[1.i] \leftarrow\u
Y[1.i] \ \mathrm{IID}  {\sim}  \operatorname{bin}(1,0.28)$}&Sc.~6\\
false&\multicolumn{6}{l}{$\u y[2.i] \leftarrow\u Y[2.i] \ \mathrm{IID}   {\sim
}  \operatorname{bin}(1,0.50),  \operatorname{corr}(\u Y[1.i],\u Y[2.i]) = 0.36,   i=1,\ldots,
100$}&\\[3pt]
Randomization\tabnoteref{ttttt2} &48.8 &NA& 58.8&58.8&48.8&60.3 \\
Process &51.1& NA& 60.1&60.1&51.1&60.3 \\
\hline
\end{tabular*}
\tabnotetext[]{}{Table entries give the rejection rates (as a percent) for the 1000 simulations.}
\tabnotetext[a]{ttttt1}{For this simulation, the permutation and Fisher randomization test
results are numerically identical.}
\tabnotetext[b]{ttttt2}{The fixed $\u y$ is such that $\u y[1.\u P] \neq \u y[2.\u P]$,
  $\overline y[1.\u P] = 24/100$,   $\overline y[2.\u P] = 45/100$,
and $\operatorname{corr}(\u y[1.\u P], \u y[2.\u P]) = 0.386$.}
\end{table*}

The simulation results in Tables~\ref{tab3}--\ref{tab6} give us a glimpse at the
operating characteristics of the tests for a variety of scenarios,
labeled ``Sc.~\#.'' The following summary focuses on
comparisons between the Fisher and Neyman randomization tests, but the
table entries afford broader comparisons.

For small $n_1, n_2$, when $\u y[1.s_j] - \u y[2.s_j] = \mathrm{constant}$, the
Neyman randomization test tends to be just a bit anti-conservative for
testing $H_0^{RAs}$; that is, the actual size appears to be a little
larger than the nominal size (see scenarios 1, 2, and 6 of Table~\ref{tab3}). This
anti-conservativeness presumably stems from the fact that the Neyman
estimator of the variance, $\operatorname{var}(D_3|\u S= \u s)$,\vadjust{\goodbreak} tends to be slightly
biased on the low side when
$\u y[1.s_j] - \u y[2.s_j] = \mathrm{constant}$.
For larger $n_1, n_2$, this anti-conservativeness disappears (scenarios
1, 2, and 6 of Table~\ref{tab5}).

When the differences $\u y[1.s_j] - \u y[2.s_j]$ are highly variable,
the Neyman randomization test tends to be a bit conservative for
testing $H_0^{RAs}$, although not as
conservative as the Fisher randomization test (scenarios 4 and 7 in
Tables~\ref{tab3} and \ref{tab5}). This
conservativeness presumably stems from the fact that the Neyman
estimator of the variance, $\operatorname{var}(D_3|\u S= \u s)$, tends to be biased on
the high side when
$\u y[1.s_j] - \u y[2.s_j]$ are highly variable (see \citett{Gad01}).

For small $n_1, n_2$, the Normal approximation to the Neyman test
statistic can be unreasonable when there are extreme outliers present
(scenario 3 of Table~\ref{tab3}). With larger $n_1, n_2$,
the Normal approximations become more reasonable in the presence of
extreme outliers (scenario~3 of Table~\ref{tab5}).

In all of the simulation scenarios, the Neyman randomization test had
higher power than the Fisher randomization test (see Tables~\ref{tab4} and \ref{tab6}),
especially when $n_1, n_2$ are smaller (see Table~\ref{tab4}). Of course,
power comparisons are most useful when both tests have the same size.
Because neither of these tests has size exactly equal to the nominal
0.05, these power comparisons should be considered carefully.
In particular, in head-to-head comparisons, the Fisher test is at a
disadvantage because its actual size is guaranteed to be no larger than
0.05; the Neyman test has size that is only approximately equal to, and
can exceed, the nominal 0.05.

On the basis of this limited simulation study, we recommend that
practitioners at least think seriously about using the Neyman
randomization test as an alternative to the Fisher randomization test,
especially when $n_1, n_2$ are moderate, say, at least 10, and when
there are no extreme outliers.

\section{Discussion}\label{sec8}

This paper used concepts from the rich literatures on causal analysis
and finite-population sampling theory to clear up some of the confusion
that exists about tests of the no-treatment-effect hypothesis in the
randomized comparative experiment setting. Our approach lends itself to
explicit specifications of the candidate no-treatment-effects
hypotheses and targets of inference. We clearly distinguished between
three main inference approaches: process-based, randomization-based,
and selection-based. The commonly-used permutation test, Wilcoxon rank
sum test, and two-sample $t$ tests are examples of process-based
approaches. Examples of randomization-based approaches include the
commonly-used Fisher randomization test and the less commonly-used
Neyman randomization test. We also described a Neyman selection test. A
small-scale empirical comparison of these different tests was carried
out. On the basis of the simulation results, we recommend that
practitioners consider using the Neyman randomization test in certain scenarios.

In our description of the process-based approach, we focused on testing
hypotheses about the distribution of $\u Y$. More generally, the
process-based approach can be used to both estimate, or test hypotheses
about, characteristics of the distribution of $\u Y$ \emph{and}
predict/estimate the unobserved values $\u y[-\u t.\u s]$. Here, $\u
y[-A]$ is the collection of all $2N$ components of $\u y$ excluding
those with subscripts in the set $A$. A look back at the assumptions
$A_1$--$A_7$ shows that we did not have to specify a model for the
joint distribution of $\u Y$ to carry out a test of no treatment
effect. We only assumed independence across units and modeled the marginal
distributions of $\u Y[1.i]$ and $\u Y[2.i]$. In contrast, the
prediction of unobserved values generally requires a model for the
joint distribution of $\u Y$, equivalently, a model for $(\u Y[\u t.\u
s],  \u Y[-\u t.\u s])$, the ``$(Y_{\mathrm{obs}}, Y_{\mathrm{mis}})$'' of Rubin (e.g.,
\citeyear{Rub05}). Rubin advocates using a Bayesian approach to process-based
prediction of $\u y[-\u t.\u s]$.

This paper restricted attention to inferences about one population or
sample, under two scenarios corresponding to two treatments. Owing to
randomization, we were able to compare these two treatment scenarios;
for example, see equation (\ref{ED}). Comparing two populations of
distinct units is a qualitatively different inference problem. However,
similar notation and model structures can be used to study this problem
as well.
Interestingly, in this two population setting, Fisher randomization
tests, as described herein, are generally not applicable. In contrast,
the other tests described in this paper, including the Neyman selection
test, are applicable.

The notation and model structure introduced in this paper can be
directly applied in more general settings where nonuniform or
constrained randomization is used or where there are more than two
treatments being compared; see, for example, the descriptions in Sutter
et al. (\citeyear{SutZysKem63}), \citet{Kem77}, and \citet{Bai81}. There are
extensions in other directions. For example, rather than testing
hypotheses, the ideas introduced in this paper show promise for
confidence interval estimation. More work in this direction will be forthcoming.

In the binary response, comparative experiment setting, \emph{Fisher's
exact test} for $2\times2$ tables (see \citett{Agr02}, page~91) is equivalent
to the Fisher randomization test of $H_0^{RUs}$ when $\u T \perp\!\!\!
\perp\u Y|\u S$ and $\u T|(\u S = \u s)$ have a uniform distribution
as in (\ref{uniformT}); recall that $H_0^{RUs}$ states that
the binary response values satisfy $\u y[1.s_j] = \u y[2.s_j],  j=1,\ldots,n$.
Fisher's exact test is also numerically equivalent to
the process-based
permutation test of $H_0^{DUP}$ when $(\u S, \u T) \perp\!\!\!\perp\u
Y$ and $\u Y[t.i]  \mathrm{indep}  \sim \operatorname{bin}(1, \pi_t)$; here $H_0^{DUP}$ is
equivalent to $\pi_1 = \pi_2$. In fact, in the simulation (scenarios 6
and 7 of Tables~\ref{tab3} and \ref{tab5}, and scenario 6 of Tables~\ref{tab4} and \ref{tab6}), because of
the uniform randomization distribution, we were able to use the R code
for Fisher's exact test, {\tt fisher.test $\{$stat$\}$}, to compute the
exact values of the Fisher randomization and permutation $p$-values.
On a related note, we point out that the Neyman randomization test is
also available for testing
the no-treatment-effect hypothesis $H_0^{RAs}$: $\overline y[1.\u s] =
\overline y[2.\u s]$ in $2\times2$ tables. This paper's simulation
results suggest that when the randomization distribution is uniform as
in (\ref{uniformT}), this Neyman randomization test for $2\times2$
tables may be somewhat more powerful than Fisher's exact test.


\section*{Acknowledgments}
Supported in part by NSF Grant SES-1059955.



\end{document}